\documentclass[11pt]{article}
\usepackage{graphicx,ifthen,color,algorithm,palatino}
\usepackage{amsmath,amsthm,amssymb,amsfonts,verbatim}
\usepackage{mathrsfs,accents,setspace}
\newboolean{ShowFigures}
\newboolean{UnBlinded}
\newboolean{DoubleSpaced}
\setboolean{ShowFigures}{true}
\setboolean{UnBlinded}{true}
\setboolean{DoubleSpaced}{false}
\def\myand{\&\ }
\def\ba{\boldsymbol{a}}
\def\bb{\boldsymbol{b}}
\def\bc{\boldsymbol{c}}

\def\be{\boldsymbol{e}}

\def\bs{\boldsymbol{s}}

\def\bu{\boldsymbol{u}}
\def\bv{\boldsymbol{v}}

\def\bx{\boldsymbol{x}}
\def\by{\boldsymbol{y}}
\def\bz{\boldsymbol{z}}
\def\bA{\boldsymbol{A}}
\def\vecof{\mbox{vec}}
\def\vech{\mbox{vech}}
\def\bD{\boldsymbol{D}}
\def\bR{\boldsymbol{R}}
\def\quarter{{\textstyle{1\over4}}}
\def\bB{\boldsymbol{B}}
\def\thickarrow{\longleftarrow}
\def\ahat{{\widehat a}}
\def\buhat{{\widehat \bu}}
\def\btheta{\boldsymbol{\theta}}
\def\bU{\boldsymbol{U}}
\def\blambda{\boldsymbol{\lambda}}
\def\bomega{\boldsymbol{\omega}}
\def\Hess{{\sf H}}
\def\Psc{{\mathcal P}}
\def\ptilde{{\widetilde p}}
\def\bDelta{\boldsymbol{\Delta}}
\def\bI{\boldsymbol{I}}
\def\sumim{\sum_{i=1}^m}
\def\infint{\int_{-\infty}^{\infty}}
\def\bib{\vskip12pt\par\noindent\hangindent=1 true cm\hangafter=1}
\def\balpha{\boldsymbol{\alpha}}
\def\Cov{\mbox{Cov}}
\def\bZ{\boldsymbol{Z}}
\def\Ksc{{\mathcal K}}
\def\Diff{{\sf D}}
\def\bmu{\boldsymbol{\mu}}
\def\Msc{{\mathcal M}}
\def\diagonal{\mbox{diagonal}}
\def\diag{\mbox{diag}}
\def\bdeta{\boldsymbol{\eta}}
\def\argmindum{\mathop{\mbox{argmin}}}
\def\argmin#1{\argmindum_{#1}}
\def\jump{\vskip3mm\noindent}
\def\bzero{\boldsymbol{0}}

\def\bSigma{\boldsymbol{\Sigma}}
\def\expit{\mbox{expit}}
\def\real{{\mathbb R}}
\def\smhalf{{\textstyle{\frac{1}{2}}}}
\def\simind{\stackrel{{\tiny \mbox{ind.}}}{\sim}}
\def\bbeta{\boldsymbol{\beta}}
\def\bdetaSUBpyijTOui{\biggerbdeta_{\mbox{\footnotesize$p(y_{ij}|\bu_i;\bbeta)\to\bu_i$}}}
\def\bdetaSUBpyijdTOui{\biggerbdeta_{\mbox{\footnotesize$p(y_{ij'}|\bu_i;\bbeta)\to\bu_i$}}}
\def\bdetaSUBuiTOpyijABV{\bdeta^{\otimes}}
\def\bdetaSUBuiTOpyij{\biggerbdeta_{\mbox{\footnotesize$\bu_i\to p(y_{ij}|\bu_i;\bbeta)$}}}
\def\mSUBuiTOpyij{
\biggerm_{\mbox{\footnotesize$\bu_i\to p(y_{ij}|\bu_i;\bbeta)$}}(\bu_i)}
\def\mSUBpyijTOui{
\biggerm_{\mbox{\footnotesize$p(y_{ij}|\bu_i;\bbeta)\to\bu_i$}}(\bu_i)}
\def\mSUBpyijdTOui{
\biggerm_{\mbox{\footnotesize$p(y_{ij'}|\bu_i;\bbeta)\to\bu_i$}}(\bu_i)}
\def\bdetaSUBSigma{\biggerbdeta_{\mbox{\scriptsize$\bSigma$}}}
\def\mSUBuiTOpui{
\biggerm_{\mbox{\footnotesize$\bu_i\to p(\bu_i;\bSigma)$}}(\bu_i)}
\def\bdetaSUBuiTOpui{
\biggerbdeta_{\mbox{\footnotesize$\bu_i\to p(\bu_i;\bSigma)$}}}
\def\bdetaSUBpuiTOui{\biggerbdeta_{\mbox{\footnotesize$p(\bu_i;\bSigma)\to\bu_i$}}}
\def\biggerm{\mbox{\large $m$}}
\def\biggerbdeta{\mbox{\large $\bold{\eta}$}}
\def\mSUBpchkyijTOui{
\biggerm_{\mbox{\footnotesize$\pcheck(y_{ij}|\bu_i;\bbeta)\to\bu_i$}}(\bu_i)}
\def\mSUBpyijTOui{
\biggerm_{\mbox{\footnotesize$p(y_{ij}|\bu_i;\bbeta)\to\bu_i$}}(\bu_i)}
\def\mSUBuiTOpchkyij{
\biggerm_{\mbox{\footnotesize$\bu_i\to\pcheck(y_{ij}|\bu_i;\bbeta)$}}(\bu_i)}
\def\mSUBuiTOpyij{
\biggerm_{\mbox{\footnotesize$\bu_i\to p(y_{ij}|\bu_i;\bbeta)$}}(\bu_i)}

\def\mSUBpuiTOui{
\biggerm_{\mbox{\footnotesize$p(\bu_i;\bSigma)\to\bu_i$}}(\bu_i)}
\def\mSUBuiTOpui{
\biggerm_{\mbox{\footnotesize$\bu_i\to p(\bu_i;\bSigma)$}}(\bu_i)}
\def\etaSUBpyijTOui{
\biggerbdeta_{\mbox{\bu_i\to\footnotesize$p(y_{ij}|\bu_i;\bbeta)$}}}
\def\etaSUBuiTOpyij{
\biggerbdeta_{\mbox{\footnotesize$\bu_i\to p(y_{ij}|\bu_i;\bbeta)$}}}

\def\etaSUBpyiTOui{
\biggerbdeta_{\mbox{\footnotesize$p(\by_i|\bu_i;\bbeta)\to\bu_i$}}}
\def\etaSUBpyijTOui{
\biggerbdeta_{\mbox{\footnotesize$p(y_{ij}|\bu_i;\bbeta)\to\bu_i$}}}
\def\etaSUBpuiTOui{
\biggerbdeta_{\mbox{\footnotesize$p(\bu_i;\bSigma)\to\bu_i$}}}
\def\buC{\bu}
\def\buCdash{\bu'}
\def\buLone{\bu^{\mbox{\tiny L1}}}
\def\buLtwo{\bu^{\mbox{\tiny L2}}}
\def\buLoneLtwoij{\left[\begin{array}{c}\buLone_i\\
\buLtwo_{ij}\end{array}\right]}
\def\bubuDashiidj{\left[\begin{array}{c}\bu_i\\
\bu'_{i'}\end{array}\right]}

\def\bSigmaLone{\bSigma^{\mbox{\tiny L1}}}
\def\bSigmaLtwo{\bSigma^{\mbox{\tiny L2}}}

\def\pcheck{\check{p}}
\def\endproof{\rightline{\rule[.2ex]{1ex}{1.5ex}}}
\def\bUbSigma{\bU_{\mbox{\scriptsize${\bSigma}$}}}
\def\bUbtheta{\bU_{\btheta}}
\def\vecbd{\mbox{vecbd}}
\def\blambdabtheta{\blambda_{\btheta}}
\def\blambdabSigma{\blambda_{\mbox{\scriptsize${\bSigma}$}}}
\def\UsubSigma{\bU_{\mbox{\tiny$\bSigma$}}}
\def\lambdaSubSigma{\blambda_{\mbox{\tiny$\bSigma$}}}
\def\bbetaTrue{\bbeta_{\mbox{\tiny true}}}
\def\bSigmaTrue{\bSigma_{\mbox{\tiny true}}}
\def\sigsqTrue{\sigma^2_{\mbox{\tiny true}}}
\def\UsubLogSigma{\bU_{\mbox{\tiny${\widehat\bthetaEP}$}}}
\def\lambdaSubLogSigma{\blambda_{\mbox{\tiny${\widehat\bthetaEP}$}}}
\def\bxF{\bx^{\mbox{\tiny F}}}
\def\bxR{\bx^{\mbox{\tiny R}}}
\def\bxRdash{\bx^{\mbox{\tiny R$\prime$}}}
\def\bxRone{\bx^{\mbox{\tiny R1}}}
\def\bxRtwo{\bx^{\mbox{\tiny R2}}}
\def\dF{d^{\mbox{\tiny F}}}
\def\dR{d^{\mbox{\tiny R}}}
\def\dRdash{d^{\mbox{\tiny R$\prime$}}}
\def\dRone{d^{\mbox{\tiny R1}}}
\def\dRtwo{d^{\mbox{\tiny R2}}}
\def\fUN{f_{\mbox{\tiny UN}}}
\def\fInput{f_{\mbox{\tiny input}}}
\def\bdetaEP{\underaccent{\sim}{\bdeta}}
\def\buEP{\underaccent{\sim}{\bu}}
\def\BPEP{\underaccent{\sim}{\mbox{BP}}}
\def\CovEP{\underaccent{\sim}{\mbox{Cov}}}
\def\pEP{\underaccent{\sim}{p}}
\def\ellEP{\underaccent{\sim}{\ell}}

\def\bomegaEP{\underaccent{\sim}{\bomega}}
\def\bbetaEP{\underaccent{\sim}{\bbeta}}
\def\bSigmaEP{\underaccent{\sim}{\bSigma}}
\def\bthetaEP{\underaccent{\sim}{\btheta}}
\def\proj{\mbox{proj}}
\def\bmuInput{\bmu^{\mbox{\tiny input}}}
\def\bSigmaInput{\bSigma^{\mbox{\tiny input}}}
\def\bdetaInput{\bdeta^{\mbox{\tiny{input}}}}
\def\bdetaOneInput{\bdeta_1^{\mbox{\tiny{input}}}}
\def\bdetaTwoInput{\bdeta_2^{\mbox{\tiny{input}}}}
\def\Cprobit{C_{\mbox{\tiny probit}}}
\def\Kprobit{K_{\mbox{\tiny probit}}}

\begin{document}
\ifthenelse{\boolean{DoubleSpaced}}{\setstretch{1.5}}{}

\thispagestyle{empty}

\centerline{\Large\bf Fast and Accurate Binary Response Mixed Model}
\vskip2mm
\centerline{\Large\bf Analysis via Expectation Propagation}
\vskip7mm
\centerline{\normalsize\sc By P. Hall$\null^1$, I.M. Johnstone$\null^2$, 
J.T. Ormerod$\null^3$, M.P. Wand$\null^4$ and J.C.F. Yu$\null^4$}
\vskip5mm
\centerline{\textit{$\null^1$University of Melbourne, 
$\null^2$Stanford University, $\null^3$University of Sydney}}
\vskip1mm
\centerline{\textit{and $\null^4$University of Technology Sydney}}
\vskip6mm
\centerline{22nd May, 2018}

\vskip6mm

\centerline{\large\bf Abstract}
\vskip2mm

Expectation propagation is a general prescription for approximation of integrals
in statistical inference problems. Its literature is mainly concerned
with Bayesian inference scenarios. However, expectation propagation can also be
used to approximate integrals arising in \emph{frequentist} statistical inference.
We focus on likelihood-based inference for binary response mixed models and show that
fast and accurate quadrature-free inference can be realized for the 
probit link case with multivariate random effects and higher levels of nesting. 
The approach is supported by asymptotic theory in which expectation propagation is seen to 
provide consistent estimation of the exact likelihood surface. Numerical 
studies reveal the availability of fast, highly accurate and scalable 
methodology for binary mixed model analysis.


\section{Introduction}\label{sec:intro}

Binary response mixed model-based data analysis is ubiquitous in many 
areas of application, with examples such as analysis of biomedical longitudinal data
(e.g.\ Diggle \textit{et al.}, 2002),  social science multilevel data (e.g.\ Goldstein, 2010),
small area survey data (e.g.\ Rao \myand Molina, 2015) 
and economic panel data (e.g.\ Baltagi, 2013).
The standard approach for likelihood-based inference in
the presence of multivariate random effects is Laplace approximation, which
is well-known to be inconsistent and prone to inferential inaccuracy.
Our main contribution is to overcome this problem using expectation propagation.
The new approach possesses speed and scalability on par with that of Laplace
approximation, but is provably consistent and demonstrably very accurate.
Bayesian approaches and Monte Carlo methods offer another route to accurate
inference for binary response mixed models (e.g. Gelman \myand Hill, 2007). 
However, speed and scalability issues aside, frequentist inference is
the dominant approach in many areas in which mixed models are used.
Henceforth, we focus on frequentist binary mixed model analysis.

The main obstacle for likelihood-based inference for binary mixed models
is the presence of irreducible integrals. For grouped data with one level
of nesting, the dimension of the integrals matches the number of random
effects. The two most common approaches to dealing with these integrals
are (1) quadrature and (2) Laplace approximation. For example, in the
\textsf{R} computing environment (R Core Team, 2018) the function
\texttt{glmer()} in the package \textsf{lme4} (Bates \textit{et al.}, 2015) 
supports both adaptive Gauss-Hermite quadrature and Laplace approximation
for univariate random effects. For multivariate random effects only Laplace
approximation is supported by \texttt{glmer()}, presumably because of
the inherent difficulties of higher dimensional quadrature. Laplace
approximation eschews multivariate integration via quadratic approximation
of the log-integrand. However, the resultant approximate inference is
well-known to be inaccurate, often to an unacceptable degree, in
binary mixed models (e.g.\ McCulloch \textit{et al.}, Section 14.4).
An embellishment of Laplace approximation, known as integrated nested Laplace
approximation (Rue, Martino \myand Chopin, 2009), has been successful in
various Bayesian inference contexts.

Expectation propagation (e.g.\ Minka, 2001) is general prescription for
approximation of integrals that arise in statistical inference problems.
Most of its literature is within the realm of Computer Science and, 
in particular, geared towards approximate inference for Bayesian graphical 
models (e.g.\ Chapter 10, Bishop, 2006). A major contribution of this
article is transferral of expectation propagation methodology 
to frequentist statistical inference. In principle, our approach
applies to any generalized linear mixed model situation. However, 
expectation propagation for binary response mixed model analysis
has some especially attractive features and therefore we focus
on this class of models. In the special case of probit mixed models,
the expectation propagation approximation to the log-likelihood
is exact regardless of the dimension of the random effects.
This leads to a new practical alternative to multivariate quadrature.
Moreover, asymptotic theory reveals that expectation propagation
provides consistent approximation of the exact likelihood surface.
This implies very good inferential accuracy of expectation propagation,
and is supported by our simulation results. We are not aware of
any other quadrature-free approaches to generalized mixed model
analysis that has such a strong theoretical underpinning.

To facilitate widespread use of the new approach, a new package in 
the \textsf{R} language (R Core Team, 2018) has been launched. 
The package, \textsf{glmmEP} (Wand \myand Yu, 2018), uses a 
low-level language implementation of expectation propagation
for speedy approximate likelihood-based inference and scales
well to large sample sizes.

Binary response mixed models, and their inherent computational
challenges, are summarized in Section \ref{sec:binMixMod}
The expectation propagation approach to fitting and approximate
inference, with special attention given to the quadrature-free
probit link situation, is given in Section \ref{sec:EPmain}.
Section \ref{sec:numerical} presents the results of numerical
studies for both simulated and real data, and shows expectation
propagation to be of great practical value as a fast, high
quality approximation that scales well to big data and big model
situations. Theoretical considerations are summarised in
Section \ref{sec:theorConsid}. Higher level and
random effects extensions are touched upon
in Section \ref{sec:highLevCrossed}. Lastly, we briefly
discuss transferral of new approach to other generalized
linear mixed model settings in Section \ref{sec:otherGLMMs}.

\section{Binary Response Mixed Models}\label{sec:binMixMod}


Binary mixed models for grouped data with one level of nesting and Gaussian
random effects has the general form
\begin{equation}
y_{ij}|\bu_i\simind\mbox{Bernoulli}\big(F\big(\bbeta^T\bxF_{ij}+\bu_i^T\bxR_{ij}\big)\big),
\quad \bu_i\simind N(\bzero,\bSigma),\quad 1\le i\le m,\quad 1\le j\le n_i
\label{eq:probitMixMod}
\end{equation}
where $F$, the \emph{inverse link}, 
is a pre-specified cumulative distribution function and $y_{ij}$ is the $j$th response for 
the $i$th group, where number of groups is $m$ and the number of responses 
measurements within the $i$th group is $n_i$. Also, $\bxF_{ij}$ is a 
$\dF\times1$ vector of predictors corresponding to $y_{ij}$,
modeled as having fixed effects with coefficient vector $\bbeta$.
Similarly, $\bxR_{ij}$ is a $\dR\times1$ vector of predictors 
modeled as having random effects with coefficient vectors $\bu_i$,\ $1\le i\le m$.
Typically, $\bxR_{ij}$ is a sub-vector of $\bxF_{ij}$. It is also very common
for each of  $\bxR_{ij}$ and $\bxF_{ij}$ to have first entry equal to $1$, 
corresponding to fixed and random intercepts.
The random effects covariance matrix $\bSigma$ has dimension $\dR\times\dR$.

By far, the most common choices for $F$ are 
$$F=\left\{\begin{array}{ll}
\expit\quad &\mbox{for logistic mixed models}\\[1ex]
\Phi\quad &\mbox{for probit mixed models}\\
\end{array}  
\right.
$$
where $\expit(x)\equiv 1/(1+e^{-x})$ and $\Phi$ is the cumulative distribution 
function of the $N(0,1)$ distribution.

Despite the simple form of (\ref{eq:probitMixMod}), likelihood-based inference
for the parameters $\bbeta$ and $\bSigma$ and best prediction of the random
effects $\bu_i$ is very numerically challenging. Assuming that $F(x)+F(-x)=1$,
as is the case for the logistic and probit cases, the log-likelihood is 
\begin{equation}
\ell(\bbeta,\bSigma)=\sum_{i=1}^m\log
\int_{\real^{\dR}}\left\{\prod_{j=1}^{n_i}F\big((2y_{ij}-1)(\bbeta^T\bxF_{ij}+\bu^T\bxR_{ij})\big)
\right\}|2\pi\bSigma|^{-1/2}\exp(-\smhalf\bu^T\bSigma^{-1}\bu)\,d\bu
\label{eq:logLikFirst}
\end{equation}
and the best predictor of $\bu_i$ is
$$\mbox{BP}(\bu_i)=
\frac{\int_{\real^{\dR}}\bu\left\{\prod_{j=1}^{n_i}F\big((2y_{ij}-1)(\bbeta^T\bxF_{ij}+\bu^T\bxR_{ij})\big)
\right\}\exp(-\smhalf\bu^T\bSigma^{-1}\bu)\,d\bu         }
{\int_{\real^{\dR}}\left\{\prod_{j=1}^{n_i}F\big((2y_{ij}-1)(\bbeta^T\bxF_{ij}+\bu^T\bxR_{ij})\big)
\right\}\exp(-\smhalf\bu^T\bSigma^{-1}\bu)\,d\bu},\quad 1\le i\le m.
$$
The $\dR$-dimensional integrals in the $\ell(\bbeta,\bSigma)$ and $\mbox{BP}(\bu_i)$ 
expressions cannot be reduced further and multivariate numerical integration must be
called upon for their evaluation. In addition, $\ell(\bbeta,\bSigma)$ has to be
maximized over $\{\dF+\smhalf\,\dR(\dR+1)\}$-dimensional space to obtain maximum 
likelihood estimates. Lastly, there is the problem of obtaining approximate 
confidence intervals for the entries of $\bbeta$ and $\bSigma$ and 
approximate prediction intervals for the entries of $\bu_i$.

\section{Expectation Propagation Likelihood Approximation}\label{sec:EPmain}

We will first explain expectation propagation for approximation
of the log-likelihood $\ell(\bbeta,\bSigma)$. Approximation
of $\mbox{BP}(\bu_i)$ follows relatively quickly. 
First note that $\ell(\bbeta,\bSigma)=\sum_{i=1}^m\ell_i(\bbeta,\bSigma)$
where
$$\ell_i(\bbeta,\bSigma)\equiv \log
\int_{\real^{\dR}}\left\{\prod_{j=1}^{n_i}F\big((2y_{ij}-1)(\bbeta^T\bxF_{ij}+\bu^T\bxR_{ij})\big)
\right\}|2\pi\bSigma|^{-1/2}\exp(-\smhalf\bu^T\bSigma^{-1}\bu)\,d\bu.$$
Each of the $\ell_i(\bbeta,\bSigma)$ are approximated individually
and then summed to approximate $\ell(\bbeta,\bSigma)$
The essence is of the approximation of $\ell_i(\bbeta,\bSigma)$ 
is replacement of each 
$$F\big((2y_{ij}-1)(\bbeta^T\bxF_{ij}+\bu^T\bxR_{ij})\big),\quad 1\le j\le n_i,$$
by an unnormalized Multivariate Normal density function, chosen according to
an appropriate minimum Kullback-Leibler divergence criterion. 
The resultant integrand is then proportional to a product of Multivariate Normal
density functions and admits an explicit form. The number
approximating density functions of the same order of magnitude and,
together with the properties of minimum Kullback-Leibler divergence,
leads to accurate and statistically consistent approximation of $\ell(\bbeta,\bSigma)$.
In probit case, where $F=\Phi$, the minimum Kullback-Leibler divergence steps
are explicit.  This leads to accurate approximation of $\ell(\bbeta,\bSigma)$
without the need for any numerical integration -- just some fixed-point iteration.
The expectation propagation-approximate log-likelihood, which we denote by 
$\ellEP(\bbeta,\bSigma)$, can be evaluated quite rapidly and maximized using
established derivative-free methods such as the Nelder-Mead algorithm (Nelder \myand Mead, 1965)
or quasi-Newton optimization methods such as the Broyden-Fletcher-Goldfarb-Shanno
approach with numerical derivatives. The latter also facilitates Hessian 
matrix approximation at the maximum, which can be used to construct 
approximate confidence intervals.

We now provide the details, with subsections on each of Kullback-Leibler
projection onto unnormalized Multivariate Normal density functions, message
passing formulation for organizing the required versions of these projections 
and quasi-Newton-based approximate inference.
The upcoming subsections require some specialized matrix notation.
If $\bA$ is $d\times d$ matrix then $\vecof(\bA)$ is
the $d^2\times 1$ vector obtained by stacking the columns
of $\bA$ underneath each other in order from left to right.
Also, $\vech(\bA)$ is $\smhalf\,d(d+1)+1$ vector defined
similarly to $\vecof(\bA)$ but only involving entries
on and below the diagonal. 
The \emph{duplication matrix of order $d$}, denoted by $\bD_d$, 
is the unique $d^2\times\smhalf d(d+1)$ matrix of zeros and ones
such that 
$$\bD_d\,\vech(\bA)=\vecof(\bA)\quad\mbox{for}\quad\bA=\bA^T.$$
The Moore-Penrose inverse of $\bD_d$ is 
$$\bD_d^{+}\equiv(\bD_d^T\bD_d)^{-1}\bD_d^T.$$

\subsection{Projection onto Unnormalized Multivariate Normal Density Functions}\label{sec:projUNN}

Let $L_1(\real^d)$ denote the set of absolutely integrable functions on $\real^d$.
For $f_1,f_2\in L_1(\real^d)$ such that $f_1,f_2\ge0$, the Kullback-Leibler
divergence of $f_2$ from $f_1$ is
\begin{equation}
\mbox{KL}(f_1\Vert f_2)=\int_{\real^d}
\big[f_1(\bx)\log\{f_1(\bx)/f_2(\bx)\}+f_2(\bx)-f_1(\bx)\big]\,d\bx
\label{eq:KLdefn}
\end{equation}
(e.g. Minka, 2005). In the special case where $f_1$ and $f_2$ are density 
functions the right-hand side of (\ref{eq:KLdefn}) reduces to the more
common Kullback-Leibler divergence expression. However, we require this
more general form that caters for \emph{unnormalized} density functions.

Now consider the family of functions on $\real^d$ of the form
\begin{equation}
\fUN(\bx)\equiv\exp\left\{
\left[
\begin{array}{c}        
1\\
\bx\\
\vech(\bx\bx^T)
\end{array}
\right]^T      
\left[
\begin{array}{c}        
\eta_0\\
\bdeta_1\\
\bdeta_2
\end{array}
\right]
 \right\}
\label{eq:fUN}
\end{equation}
where $\eta_0\in\real$, $\bdeta_1$ is a $d\times1$ vector and 
$\bdeta_2$ is a $\smhalf\,d(d+1)\times 1$ vector restricted in
such a way that $\fUN\in L_1(\real^d)$. Then (\ref{eq:fUN})
is the family of unnormalized Multivariate Normal density 
functions written in exponential family form with natural
parameters $\eta_0$, $\bdeta_1$ and $\bdeta_2$.

Expectation propagation for generalized linear mixed models
with Gaussian random effects has the following notion
at its core:
\begin{equation}
\mbox{given $\fInput\in L_1(\real^d)$, determine the
$\eta_0$, $\bdeta_1$ and $\bdeta_2$ that minimizes $\mbox{KL}(\fInput\Vert\,\fUN)$.}
\label{eq:KLprob}
\end{equation}

\noindent
The solution is termed the (Kullback-Leibler) projection onto the family
of Multivariate Normal density functions and we write
$$\proj[\fInput](\bx)\equiv
\exp\left\{
\left[
\begin{array}{c}        
1\\
\bx\\
\vech(\bx\bx^T)
\end{array}
\right]^T      
\left[
\begin{array}{c}        
\eta_0^*\\
\bdeta_1^*\\
\bdeta_2^*
\end{array}
\right]
 \right\}
$$
where 
$$(\eta_0^*,\bdeta_1^*,\bdeta_2^*)=\argmin{(\eta_0,\bdeta_1,\bdeta_2)\in H}
\mbox{KL}\big(\fInput\Vert\,\fUN\big),
$$
with $H$ denoting the set of all allowable natural parameters. Note that
the special case of Kullback-Leibler projection onto the unnormalized
Multivariate Normal family has a simple moment-matching representation,
with $(\eta_0^*,\bdeta_1^*,\bdeta_2^*)$ being the unique vector such
that zeroth-, first- and second-order moments of $\fUN$ match those
of $\fInput$.

For the binary mixed model (\ref{eq:probitMixMod}), expectation propagation
requires repeated projection of the form 
$$\fInput(\bx)=F(c_0+\bc_1^T\bx)\exp\left\{
\left[  
\begin{array}{c}
\bx\\[1ex]
\vech(\bx\bx^T)
\end{array}
\right]^T   
\left[  
\begin{array}{c}
\bdetaOneInput\\[1ex]
\bdetaTwoInput
\end{array}
\right]
\right\}
$$
onto the unnormalized Multivariate Normal family.
An important observation is that case of probit mixed models,
$\proj[\fInput](\bx)$ has an exact solution.

Let $\zeta(x)\equiv\log\{2\Phi(x)\}$. It follows that 
$$\zeta'(x)=\phi(x)/\Phi(x)\quad\mbox{and}\quad\zeta''(x)=-\zeta'(x)\{x+\zeta'(x)\}$$
where $\phi(x)\equiv(2\pi)^{-1/2}\exp(-\smhalf\,x^2)$ is the $N(0,1)$ density function.
We are now in a position to define two algebraic functions which are fundamental
for approximate likelihood-based inference in probit mixed models based on
expectation propagation:
\jump

\noindent
{\bf Definition 1.} \textit{For primary arguments $\ba_1$ ($d\times1$) and $\ba_2$ 
($\smhalf\,d(d+1)\times1$) such that} $\vecof^{-1}(-\bD_d^{+T}\ba_2)$ \textit{is symmetric and positive definite,
and auxiliary arguments $c_0\in\real$ and $\bc_1$ ($d\times 1$) the function} 
$\Kprobit$ \textit{is given by}
$$
\Kprobit\left(\left[\begin{array}{c}   
\ba_1\\
\ba_2
\end{array}
\right];c_0,\bc_1\right)
\equiv
\left[\setlength\arraycolsep{0.5pt}{
\begin{array}{c}   
\bR_5^T(\ba_1+r_3\bc_1)\\[3ex]
\bD_d^T\vecof(\bR_5^T\bA_2)
\end{array}}
\right]
$$
\textit{with}
$$\bA_2\equiv\vecof^{-1}(\bD_d^{+T}\ba_2),\quad
r_1\equiv\sqrt{2(2-\bc_1^T\bA_2^{-1}\bc_1)},\quad
r_2\equiv(2c_0-\bc_1^T\bA_2^{-1}\ba_1\big)/r_1,
$$
$$r_3\equiv2\zeta'(r_2)/r_1,\quad
r_4\equiv -2\zeta''(r_2)/r_1^2\quad\mbox{\textit{and}}\quad
\bR_5\equiv\big(\bA_2+r_4\bc_1\bc_1^T\big)^{-1}\bA_2
$$
\jump
\textit{and the function} $A_N$ \textit{is given by}
$$A_N\left(
\left[\begin{array}{c}   
\ba_1\\
\ba_2
\end{array}
\right]\right)
\equiv -\quarter\ba_1^T\bA_2^{-1}\ba_1-\smhalf\log\Big|-2\bA_2\Big|.
$$
\textit{In addition, for primary arguments $\ba_1$,$\bb_1$ ($d\times1$) and 
$\ba_2,\bb_2$ ($\smhalf\,d(d+1)\times 1$) such that both} $\vecof^{-1}(-\bD_d^{+T}\ba_2)$ \textit{and} 
$\vecof^{-1}(-\bD_d^{+T}\bb_2)$ 
\textit{are symmetric and positive definite, and auxiliary arguments $c_0\in\real$ and $\bc_1$ ($d\times 1$),
the function} $\Cprobit$ \textit{is given by}
$$
\Cprobit\left(
\left[\begin{array}{c}   
\ba_1\\
\ba_2
\end{array}
\right],
\left[\begin{array}{c}   
\bb_1\\
\bb_2
\end{array}
\right]
;c_0,\bc_1\right)
\equiv\log\Phi(r_2)+\quarter\,\bb_1^T\bB_2^{-1}\bb_1-\quarter\,\ba_1^T\bA_2^{-1}\ba_1\\[1ex]
+\smhalf\log\{|\bB_2|/|\bA_2|\}
$$
\textit{with} $\bB_2\equiv \vecof^{-1}(\bD_d^{+T}\bb_2)$.

\jump
Inspection of Definition 1 reveals that the $\Kprobit$ and $\Cprobit$ functions
are simple functions up to evaluations of $\log(\Phi)$ and $\zeta'=\phi/\Phi$. Even though
software for $\Phi$ is widely available, direct computation of $\log(\Phi)$ and
$\zeta'$ can be unstable and software such as the function \texttt{zeta()} in
the \textsf{R} package \textsf{sn} (Azzalini, 2017) is recommended. Another option
is use of continued fraction representation and Lentz's Algorithm (e.g. Wand \myand Ormerod, 2012).

Expectation propagation for probit mixed models relies heavily upon:
\jump
{\bf Theorem 1.} \textit{If} 
$$\fInput(\bx)=\Phi(c_0+\bc_1^T\bx)\exp\left\{  
\left[\begin{array}{c}
\bx\\
\vech(\bx\bx^T)
\end{array}
\right]^T
\left[  
\begin{array}{c}
\bdetaOneInput\\[1ex]
\bdetaTwoInput
\end{array}
\right]
\right\} 
$$
\textit{then}
$$\proj[\fInput](\bx)=\exp\left\{
\left[
\begin{array}{c}
1\\
\bx\\
\vech(\bx\bx^T)
\end{array}
\right]^T   
\left[
\begin{array}{c}
\eta_0^*\\
\bdeta_1^*\\
\bdeta_2^*
\end{array}
\right]
\right\}$$
\textit{where}
$$\left[
\begin{array}{c}
\bdeta_1^*\\
\bdeta_2^*
\end{array}
\right]=\Kprobit\left(\left[
\begin{array}{c}
\bdetaOneInput\\
\bdetaTwoInput
\end{array}
\right];c_0,\bc_1\right)\quad\mbox{\textit{and}}\quad
\eta_0^*=\Cprobit\left(
\left[
\begin{array}{c}
\bdetaOneInput\\
\bdetaTwoInput
\end{array}
\right],
\left[
\begin{array}{c}
\bdeta_1^*\\
\bdeta_2^*
\end{array}
\right]
;c_0,\bc_1\right).
$$

A proof of Theorem 1 is given in Section \ref{sec:proofThmOne} of the 
online supplement.

\subsection{Message Passing Formulation}\label{sec:messPassFormul}

\begin{figure}[!ht]
\centering
{\includegraphics[width=0.65\textwidth]{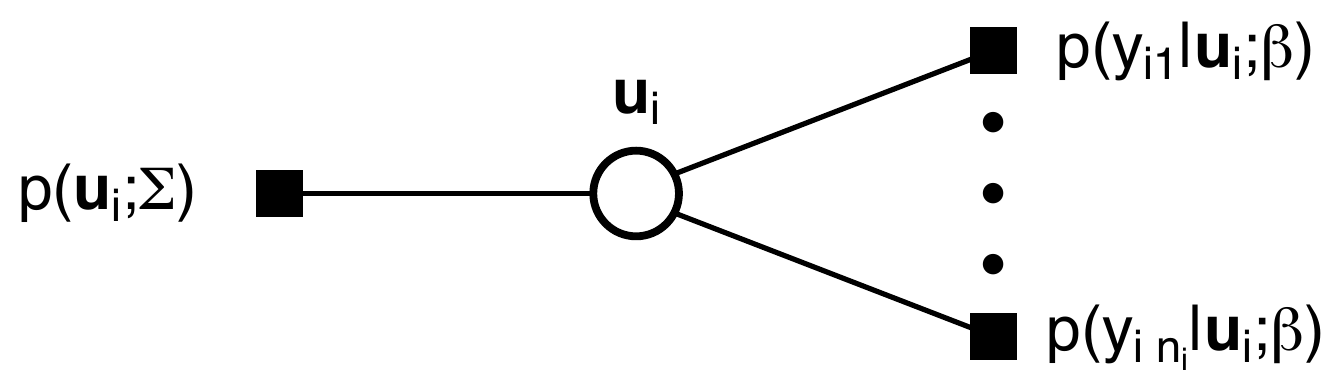}}
\caption{\it Factor graph representation of the product structure of
the integrand in (\ref{eq:elli}). The open circle corresponds to the 
random effect vector $\bu_i$ and the solid rectangles indicate factors.
Edges indicate dependence of each factor on $\bu_i$.}
\label{fig:frqMixModFacGraph} 
\end{figure}

The $i$th summand of $\ell(\bbeta,\bSigma)$ can be written as
\begin{equation}
\ell_i(\bbeta,\bSigma)=\log\int_{\real^{\dR}}\left\{\prod_{j=1}^{n_i}p(y_{ij}|\bu_i;\bbeta)\right\}
p(\bu_i;\bSigma)\,d\bu_i
\label{eq:elli}
\end{equation}
where, for $1\le j\le n_i$, 
$$p(y_{ij}|\bu_i;\bbeta)\equiv F\big((2y_{ij}-1)(\bbeta^T\bxF_{ij}+\bu_i^T\bxR_{ij})\big)
\quad\mbox{and}\quad
p(\bu_i;\bSigma)\equiv|2\pi\bSigma|^{-1/2}\exp\big(-\smhalf\bu_i^T\bSigma^{-1}\bu_i\big)
$$
are, respectively, the conditional density functions of each response given its
random effect and the density function of that random effect.
Note that product structure of the integrand in (\ref{eq:elli}) can
be represented using \emph{factor graph} shown in Figure \ref{fig:frqMixModFacGraph}.
The circle in Figure \ref{fig:frqMixModFacGraph} corresponds to the random vector
$\bu_i$ and factor graph parlance is a \emph{stochastic variable node}. The solid rectangles
correspond to each of the $n_i+1$ \emph{factors} in the (\ref{eq:elli}) integrand. 
Each of these factors depend on $\bu_i$, which is signified by an edges connecting
each factor node to the lone stochastic variable node.

Expectation propagation approximation of $\ell_i(\bbeta,\bSigma)$ involves
projection onto the unnormalized Multivariate Normal family.
Suppose that
\begin{equation}
\pEP(y_{ij}|\bu_i;\bbeta)=
\exp\left\{
\left[
\begin{array}{c}        
1\\
\bu_i\\
\vech(\bu_i\bu_i^T)
\end{array}
\right]^T      
\bdeta_{ij}
\right\},\quad 1\le j\le n_i
\label{eq:pEPexpn}
\end{equation}
are initialized to be unnormalized Multivariate Normal density 
functions in $\bu_i$. Then, for each $j=1,\ldots,n_i$, the $\bdeta_{ij}$ update
involves minimization of
\begin{equation}
\mbox{KL}\Bigg(
p(y_{ij}|\bu_i;\bbeta)
\left\{\prod_{j'\ne j}^{n_i}\pEP(y_{ij'}|\bu_i;\bbeta)\right\}p(\bu_i;\bSigma)\,
\Bigg\Vert\,
\left\{\prod_{j'=1}^{n_i}\pEP(y_{ij'}|\bu_i;\bbeta)\right\}p(\bu_i;\bSigma)\Bigg)
\label{eq:KLijth}
\end{equation}
as functions of $\bu_i$. Noting that this problem has the form (\ref{eq:KLprob}), 
Theorem 1 can be used to perform the update explicitly in the case of a probit link.
This procedure is then iterated until the $\bdeta_{ij}$s converge.

A convenient way to keep track of the updates and compartmentalize the
algebra and coding is to call upon the notion of \emph{message passing}.
Minka (2005) shows how to express expectation propagation as a message passing
algorithm in the Bayesian graphical models context, culminating in his equation (54)
and (83) update formulae. Exactly the same formulae arise here, as is
made clear in Section \ref{sec:mainAlgoDerivn} of the online supplement.
In particular, in keeping with (83) of Minka (2005), (\ref{eq:KLijth}) can be expressed as 
\begin{equation}
\mSUBpyijTOui\thickarrow
\frac{\proj\big[\mSUBuiTOpyij\,p(y_{ij}|u_i;\bbeta)\big](\bu_i)}
{\mSUBuiTOpyij},\quad 1\le j\le n_i,
\label{eq:FtoSI}
\end{equation}
where $\mSUBpyijTOui$ is the \emph{message} passed from the factor
$p(y_{ij}|\bu_i;\bbeta)$ to the stochastic node $\bu_i$ and
$\mSUBuiTOpyij$ is the message passed from $\bu_i$ back to 
$p(y_{ij}|\bu_i;\bbeta)$. The message passed from $p(\bu_i;\bSigma)$
to $\bu_i$ is 
\begin{equation}
\mSUBpuiTOui\thickarrow
\frac{\proj\big[\mSUBuiTOpui\,p(u_i;\bSigma)\big](\bu_i)}
{\mSUBuiTOpui}.
\label{eq:FtoSII}
\end{equation}
In keeping with equation (54) of Minka (2005), the stochastic node to 
factor messages are updated according to 
\begin{equation}
\mSUBuiTOpyij=\mSUBpuiTOui\left\{\prod_{j'\ne j}^{n_i}\mSUBpyijdTOui\right\},
\ 1\le j\le n_i,
\label{eq:StoFI}
\end{equation}
and
\begin{equation}
\mSUBuiTOpui=\prod_{j=1}^{n_i}\mSUBpyijTOui.
\label{eq:StoFII}
\end{equation}
As laid out at the end of Section 6 of Minka (2005), the expectation message 
passing protocol is:

\begin{minipage}[t]{132mm}
\hrule
\vskip3mm
\begin{itemize}
\item[] Initialize all factor to stochastic node messages.
\item[] Cycle until all factor to stochastic node messages converge:
\begin{itemize}
\item[]For each factor:
\begin{itemize}
\item[] Compute the messages passed to the factor using (\ref{eq:StoFI})
or (\ref{eq:StoFII}).
\item[] Compute the messages passed from the factor using (\ref{eq:FtoSI})
or (\ref{eq:FtoSII}).
\end{itemize}
\end{itemize}
\end{itemize}
\hrule
\end{minipage}

\vskip5mm
Upon convergence, the expectation propagation propagation approximation to $\ell_i(\bbeta,\bSigma)$ is
\begin{equation}
\ellEP_i(\bbeta,\bSigma)=\log\int_{\real^{\dR}}\left\{\prod_{j=1}^{n_i}
\mSUBpyijTOui\right\}\mSUBpuiTOui\,d\bu_i.
\label{eq:ellEPi}
\end{equation}
where the integrand is in keeping with the general form given by (44) of Minka \myand Winn (2008).
The success of expectation propagation hinges on the fact that each of the messages in
(\ref{eq:ellEPi}) is an unnormalized Multivariate Normal density function and the integral
over $\real^{\dR}$ can be obtained exactly as follows:
\begin{eqnarray*}
&&\int_{\real^{\dR}}\left\{\prod_{j=1}^{n_i}
\mSUBpyijTOui\right\}\mSUBpuiTOui\,d\bu_i\\[1ex]
&&\qquad\qquad=\int_{\real^{\dR}}
\left[\prod_{j=1}^{n_i}
\exp\left\{
\left[
\begin{array}{c}        
1\\
\bu_i\\
\vech(\bu_i\bu_i^T)
\end{array}
\right]^T      
\etaSUBpyijTOui\right\}\right]\\[1ex]
&&\qquad\qquad\qquad\times\exp\left\{
\left[
\begin{array}{c}        
1\\
\bu_i\\
\vech(\bu_i\bu_i^T)
\end{array}
\right]^T      
\etaSUBpuiTOui\right\}
\,d\bu_i\\[1ex]
&&\qquad\qquad=(2\pi)^{-1/2}
\exp\left\{\left(\biggerbdeta_{\bSigma}+\mbox{SUM}\{\etaSUBpyiTOui\}\right)_0\right.\\[1ex]
&&\qquad\qquad\qquad\qquad
+\left.A_N\left(\left(\biggerbdeta_{\bSigma}+\mbox{SUM}\{\etaSUBpyiTOui\}\right)_{-0}\right)\right\}
\end{eqnarray*}
where 
$$\biggerbdeta_{\bSigma}\equiv\left[\begin{array}{c}  
-\smhalf\log|2\pi\bSigma|\\[1ex]
\bzero_{\dR}\\[1ex]
-\smhalf\bD_{\dR}^T\vecof(\bSigma^{-1})
\end{array}
\right],\quad
\mbox{SUM}\{\etaSUBpyiTOui\}\equiv\displaystyle{\sum_{j=1}^{n_i}}\etaSUBpyijTOui,
$$
$A_N$ is as defined in Definition 1 and,
for an unnormalized Multivariate Normal natural parameter vector $\bdeta$,
$\bdeta_0$ denotes the first entry (the zero subscript is indicative
of the first entry being the coefficient of $1$) and $\bdeta_{-0}$ denotes
the remaining entries.

The full algorithm for expectation propagation approximation of $\ell(\bbeta,\bSigma)$
is summarized as Algorithm \ref{alg:mainAlgo}. The derivational details are given in
Section \ref{sec:mainAlgoDerivn}.

\begin{algorithm}[!th]
\begin{center}
\begin{minipage}[t]{160mm}
\hrule
\jump
\vskip2mm
\begin{small}
\begin{itemize}
\item[] Inputs: $y_{ij},\bxF_{ij},\bxR_{ij}$, $1\le i\le m$, $1\le j\le n_i$;\\[1ex]
\null$\ \ \ \ \ \ \ \ \ \ \ \ \ $
$\bbeta$ $(\dF\times 1)$, $\bSigma$ $(\dR\times\dR, \mbox{symmetric and positive definite}).$
\item[] Set constants: $\ c_{0,ij}\thickarrow(2y_{ij}-1)(\bbeta^T\bxF_{ij});\ \bc_{1,ij}\thickarrow(2y_{ij}-1)\bxR_{ij},
                       \qquad 1\le i\le m,\ 1\le j\le n_i;$
\item[]\null$\qquad\qquad\qquad\quad\etaSUBpuiTOui\thickarrow\biggerbdeta_{\bSigma}\equiv\left[\begin{array}{c}  
-\smhalf\log|2\pi\bSigma|\\[1ex]
\bzero_{\dR}\\[1ex]
-\smhalf\bD_{\dR}^T\vecof(\bSigma^{-1})
\end{array}
\right],\quad 1\le i\le m.$
\item[] For $i=1,\ldots,m$:
\begin{itemize}
\item[] Initialize: $\qquad\etaSUBpyijTOui,\quad  1\le j\le n_i$\ \ (see Section \ref{sec:sttVals} for a recommendation)
\item[] Cycle:
\begin{itemize}
\item[] $\mbox{SUM}\{\etaSUBpyiTOui\}\thickarrow\displaystyle{\sum_{j=1}^{n_i}}\etaSUBpyijTOui$\\[0.5ex]
\item[] For $j=1,\ldots,n_i$:\\[0ex]
\begin{itemize}
\item[] $\etaSUBuiTOpyij\thickarrow\etaSUBpuiTOui+\mbox{SUM}\{\etaSUBpyiTOui\}-\etaSUBpyijTOui$\\[1ex]
\item[] $\Big(\etaSUBpyijTOui\Big)_{-0}\thickarrow\Kprobit\Big(\big(\etaSUBuiTOpyij\big)_{-0};c_{0,ij},\bc_{1,ij}\Big)$
\item[] $\qquad\qquad\qquad\qquad\qquad\qquad\qquad\qquad\qquad-\big(\etaSUBuiTOpyij\big)_{-0}$\\[0ex]
\end{itemize} 
\end{itemize}
\item[] until all natural parameter vectors converge.
\item[] For $j=1,\ldots,n_i$:\\[0ex]
\begin{itemize}
\item[] $\Big(\etaSUBpyijTOui\Big)_0\thickarrow\Cprobit\Big(\big(\etaSUBuiTOpyij\big)_{-0},
\big(\etaSUBpyijTOui\big)_{-0}$
\item[]$\qquad\qquad\qquad\qquad\qquad\qquad\qquad\quad\qquad
+\big(\etaSUBuiTOpyij\big)_{-0};c_{0,ij},\bc_{1,ij}\Big)$
\end{itemize}
\item[] $\mbox{SUM}\{\etaSUBpyiTOui\}\thickarrow\displaystyle{\sum_{j=1}^{n_i}}\etaSUBpyijTOui$\\[1ex]
\end{itemize}
\item[] Output: The expectation propagation approximate log-likelihood given by
\setlength\arraycolsep{2pt}{
\begin{eqnarray*}
\ellEP(\bbeta,\bSigma)&=&\smhalf\,m\log(2\pi)+\sum_{i=1}^m\Big\{
\left(\biggerbdeta_{\bSigma}+\mbox{SUM}\{\etaSUBpyiTOui\}\right)_0\\[1ex]
&&\qquad\qquad\qquad
+A_N\Big(\left(\biggerbdeta_{\bSigma}+\mbox{SUM}\{\etaSUBpyiTOui\}\right)_{-0}\Big)
\Big\}
\end{eqnarray*}
}
\end{itemize}
\end{small}
\vskip1mm
\hrule
\end{minipage}
\end{center}
\caption{\it Expectation expectation approximation of the log-likelihood for the probit
mixed model (\ref{eq:probitMixMod}) with $F=\Phi$ via message passing on the
Figure \ref{fig:frqMixModFacGraph} factor graph.}
\label{alg:mainAlgo} 
\end{algorithm}

We have carried out extensive simulated data tests on Algorithm \ref{alg:mainAlgo}
using the starting values described in Section \ref{sec:sttVals} and found 
convergence to be rapid. Moreover, each of updates in Algorithm \ref{alg:mainAlgo}
involve explicit calculations and low-level language implementation, used in 
our \textsf{R} package \textsf{glmmEP}, affords very fast evaluation of the approximate 
log-likelihood surface. As explained in (\ref{sec:quasiNewton}), quasi-Newton methods
can be used for maximization of $\ellEP(\bbeta,\bSigma)$ and approximate
likelihood-based inference.

\subsection{Recommended Starting Values for Algorithm \ref{alg:mainAlgo}}\label{sec:sttVals}

In Section \ref{sec:sttValsDeriv} we use a Taylor series argument to justify the following
starting values for $\etaSUBpyijTOui$ in Algorithm \ref{alg:mainAlgo}:
\begin{equation}
\etaSUBpyijTOui^{\mbox{\tiny start}}\equiv
\left[
\begin{array}{c}
0\\[0.5ex]
(2y_{ij}-1)\zeta'(\ahat_{ij})\bxR_{ij}-\zeta''(\ahat_{ij})\bxR_{ij}(\bxR_{ij})^T\buhat_i\\[2ex]
\smhalf\zeta''(\ahat_{ij})\bD_{\dR}^T\vecof\big(\bxR_{ij}(\bxR_{ij})^T\big)
\end{array}
\right],\quad 1\le j\le n_i,
\label{eq:etaSttExpr}
\end{equation}
where
$$\ahat_{ij}\equiv(2y_{ij}-1)(\bbeta^T\bxF_{ij}+\buhat_i^T\bxR_{ij})$$
and $\buhat_i$ is a prediction of $\bu_i$. A convenient choice for $\buhat_i$
is that based on Laplace approximation. In the \textsf{R} computing environment
the function \texttt{glmer()} in the package \textsf{lme4} (Bates \textit{et al.}, 2015)
provides fast Laplace approximation-based predictions for the $\bu_i$.
In our numerical experiments, we found convergence of the cycle
loop of Algorithm \ref{alg:mainAlgo} to be quite rapid, with convergents of 
$$\Big(\etaSUBpyijTOui\Big)_{-0}\quad\mbox{relatively close to}\quad
\Big(\etaSUBpyijTOui^{\mbox{\tiny start}}\Big)_{-0}.$$
Therefore, we strongly recommend the starting values (\ref{eq:etaSttExpr}).

\subsection{Quasi-Newton Optimization and Approximate Inference}\label{sec:quasiNewton}

Even though Algorithm \ref{alg:mainAlgo} provides fast approximate evaluation
of the probit mixed model likelihood surface, we still need to maximize over
$(\bbeta,\bSigma)$ to obtain the expectation propagation-approximate maximum
likelihood estimators $({\widehat\bbetaEP},{\widehat\bSigmaEP})$. This is also the issue of 
approximate inference based on Fisher information theory.

Since $\ellEP(\bbeta,\bSigma)$ is defined implicitly via an iterative scheme,
differentiation for use in derivative-based optimization techniques is not straightforward.
A practical workaround involves the employment of optimization
methods such as those of the quasi-Newton variety for which 
derivatives are approximated numerically. In the \textsf{R} computing
environment the function \texttt{optim()} supports several derivative-free
optimization implementations. The \textsf{Matlab} computing environment 
(The Mathworks Incorporated, 2018) has similar capabilities via functions
such as \texttt{fminunc()}.
In the \textsf{glmmEP} package and
the examples in Section \ref{sec:numerical} we use the Broyden-Fletcher-Goldfarb-Shanno 
quasi-Newton method (Broyden, 1970; Fletcher 1970; Goldfarb, 1970; Shanno, 1970)  
with Nelder-Mead  starting values. Section 2.2.2.3 of Givens \myand Hoetig (2005) provides
a concise summary of the Broyden-Fletcher-Goldfarb-Shanno method. 

Since $\bSigma$ is constrained to be symmetric and positive definite, 
we instead perform quasi-Newton optimization over the unconstrained
parameter vector $(\bbeta,\btheta)$ where
$$\btheta\equiv\vech\big(\smhalf\log(\bSigma)\big)$$
and $\log(\bSigma)$ is the matrix logarithm of $\bSigma$
(e.g. Section 2.2 of Pinheiro \myand Bates, 2000).
Note that $\log(\bSigma)$ can be obtained using
$$\log(\bSigma)=\UsubSigma\mbox{diag}\{\log(\lambdaSubSigma)\}\UsubSigma^T\quad\mbox{where}\quad
\bSigma=\UsubSigma\mbox{diag}(\lambdaSubSigma)\UsubSigma^T
$$
is the spectral decomposition of $\bSigma$ and $\log(\lambdaSubSigma)$ denotes element-wise
evaluation of the logarithm to the entries of $\lambdaSubSigma$.
If $({\widehat\bbetaEP},{\widehat\bthetaEP})$ is the maximizer of $\ellEP$ then
the expectation propagation-approximate maximum likelihood estimate of $\bSigma$ is
$${\widehat\bSigmaEP}=\UsubLogSigma\mbox{diag}\{\exp(2\lambdaSubLogSigma)\}\UsubLogSigma^T
\quad\mbox{where}\quad\vech^{-1}({\widehat\bthetaEP})
=\UsubLogSigma\mbox{diag}(\lambdaSubLogSigma)\UsubLogSigma^T
$$
is the spectral decomposition of the $\vech^{-1}({\widehat\bthetaEP})$.
Note that $\vech^{-1}(\ba)$ is the symmetric matrix $\bA$ of appropriate
dimension such that $\vech(\bA)=\ba$.

The \texttt{optim()} function in \textsf{R} and the \texttt{fminunc()} function
in \textsf{Matlab} each have the option of computing
an approximation to the Hessian matrix at the optimum, which can be used
for approximate likelihood-based inference. 
In particular, we can use the approximate Hessian matrix to construct confidence intervals
for the entries of $\bbeta$ and the standard deviation and correlation parameters of $\bSigma$. 
The full details are given in Section \ref{sec:confIntDetails} of
the online supplement. Here we sketch the idea for the special
case of $\dR=2$, for which 
$$\bSigma=\left[\begin{array}{cc}          
\sigma_1^2 & \rho\sigma_1\sigma_2 \\
\rho\sigma_1\sigma_2 & \sigma_2^2
\end{array}\right].
$$
For confidence interval construction it is appropriate (e.g. Section 2.4 of Pinheiro \myand Bates)
to work with the parameter vector
$$\bomega\equiv\left[
\begin{array}{c}
\log(\sigma_1)\\
\log(\sigma_2)\\
\tanh^{-1}(\rho)
\end{array}
\right].
$$
Approximate $100(1-\alpha)\%$ confidence intervals for the entries of $(\bbeta,\bomega)^T$ are
\begin{equation}
\left[\begin{array}{c}       
{\widehat\bbetaEP}\\[2ex]
{\widehat\bomegaEP}
\end{array}
\right]\pm\Phi^{-1}(1-\smhalf\,\alpha)\sqrt{-\mbox{diagonal}\big(
\{\Hess\,\ellEP(\widehat{\bbetaEP},\widehat{\bomegaEP})\}^{-1}\big)}
\label{eq:Ciomega}
\end{equation}
where $\Hess\,\ellEP(\bbeta,\bomega)$ is the Hessian matrix of $\ellEP$
with respect to the $(\bbeta,\bomega)$ parameter vector.
Confidence intervals for the entries of $\bbeta$, $\sigma_1$, 
$\sigma_2$ and $\rho$ follow from standard inversion manipulations.

Note that $(\bbeta,\theta)$ is an unconstrained parametrization whilst 
$(\bbeta,\bomega)$ is a constrained parametrization. Hence, the optimization 
should be performed with respect to the former parametrization whereas
the Hessian matrix in (\ref{eq:Ciomega}) is respect to the latter parametrization.
In the examples of Section \ref{sec:numerical} and the \textsf{R} package
\textsf{glmmEP} we use the following strategy:
\begin{itemize}
\item Obtain $({\widehat\bbetaEP},{\widehat\bthetaEP})$ using \texttt{optim()}
with the $(\bbeta,\btheta)$ parametrization in the function being maximized
and the \texttt{hessian} argument set to \texttt{FALSE}.
\item Compute $({\widehat\bbetaEP},{\widehat\bomegaEP})$ and use this as
a initial value with a call to \texttt{optim()} with the 
$(\bbeta,\bomega)$ parametrization in the function being maximized
and the \texttt{hessian} argument set to \texttt{TRUE}.
\end{itemize}

Full details of confidence interval calculations for the general
multivariate random effects situation are given in Section \ref{sec:confIntDetails}
of the online supplement.

In our numerical experiments, we have found Nelder-Mead followed 
by Broyden-Fletcher-Goldfarb-Shanno optimization of expectation 
propagation approximate log-like\-li\-hood, with confidence intervals 
based on the approximate Hessian matrix, to be very effective.
In Section \ref{sec:numerical} we present simulation results
that show this strategy producing fast and accurate inference
for binary mixed models.

\subsection{Expectation Propagation Approximate Best Prediction}

The best predictors of $\bu_i$ are
$$\mbox{BP}(\bu_i)\equiv E(\bu_i|\by),\quad 1\le i\le m.$$
We now show that Algorithm \ref{alg:mainAlgo} provides, 
as by-products, straightforward empirical best predictions
of the $\bu_i$.  

Let 
\begin{equation}
{\widehat\bdetaEP}_i\equiv\biggerbdeta_{\bSigma}+\mbox{SUM}\{\etaSUBpyiTOui\}=
\left[\begin{array}{c}  
{\widehat\bdetaEP}_{i1}\\[1ex]
{\widehat\bdetaEP}_{i2}\\[1ex]
\end{array}
\right]
\label{eq:bdetaEPdfn}
\end{equation}
where $\biggerbdeta_{\bSigma}$ and $\mbox{SUM}\{\etaSUBpyiTOui\}$ are 
as in Algorithm \ref{alg:mainAlgo} with $(\bbeta,\bSigma)=(\widehat{\bbetaEP},\widehat{\bSigmaEP})$,
${\widehat\bdetaEP}_{i1}$ is the sub-vector of ${\widehat\bdetaEP}_i$ corresponding
to the first $\dR$ entries and ${\widehat\bdetaEP}_{i2}$ contains the remaining entries.
Then in Section \ref{sec:bestPredDetails} of the online supplement
we show that a suitable empirical approximation to 
$\mbox{BP}(\bu_i)$, based on the expectation propagation estimate, is
\begin{equation}
\BPEP(\bu_i)=-\smhalf\Big\{\vecof^{-1}\Big(
\bD_d^{+T}{\widehat\bdetaEP}_{i2}\Big)\Big\}^{-1}{\widehat\bdetaEP}_{i1}.
\label{eq:BPdefn}
\end{equation}
The corresponding covariance matrix empirical approximation is
\begin{equation}
\CovEP(\bu_i|\by)=-\smhalf\Big\{\vecof^{-1}\Big(
\bD_d^{+T}{\widehat\bdetaEP}_{i2}\Big)\Big\}^{-1}.
\label{eq:sampVersion}
\end{equation}
In view of equation (13.7) of McCulloch, Searle \myand Neuhaus (2008),
$\mbox{Cov}\{\BPEP(\bu_i)-\bu_i\}$ is approximated by 
$E_{\by_i}\{\CovEP(\bu_i|\by_i)\}$. Approximate prediction interval construction 
is hindered by this expectation over the sampling distribution of the 
responses. See, for example, Carlin \myand Gelfand (1991), for discussion and 
access to some of the relevant literature concerning valid prediction interval
construction in the more general empirical Bayes context.

\section{Numerical Evaluation and Illustration}\label{sec:numerical}

We now demonstrate the impressive accuracy and speed of Algorithm \ref{alg:mainAlgo}
combined with quasi-Newton methods for approximate likelihood-based inference
for probit mixed models. Firstly, we report the results of some studies involving
simulated data. Analysis of actual data is discussed later in this section.

\subsection{Simulations}

Our simulations involved (1) comparison with exact maximum likelihood for the 
$\dR=1$ situation for which quadrature is univariate, and (2) evaluation
of inferential accuracy and speed for a larger model involving bivariate
random effects.

\subsubsection{Comparison with Exact Maximum Likelihood for Univariate Random Effects}

Our first simulation study involved simulation of 1,000 datasets according
to the $\dR=1$ version of (\ref{eq:probitMixMod}) with true parameter values:
\begin{equation}
\bbetaTrue=[0,1]^T\quad\mbox{and}\quad\bSigmaTrue=\sigsqTrue=1.
\label{eq:trueParmsFirst}
\end{equation}
The sample sizes were set to $m=100$ and $n_i=2$.
The $\bxF_{ij}$ and $\bxR_{ij}$ vectors were of the form
\begin{equation}
\bxF_{ij}=[1,x_{ij}]^T\quad\mbox{and}\quad\bxR_{ij}=1
\label{eq:xFandxRsimOne}
\end{equation}
where $x_{ij}$ was generated independently from a Uniform
distribution on the unit interval. 

\begin{figure}[!ht]
\centering
{\includegraphics[width=\textwidth]{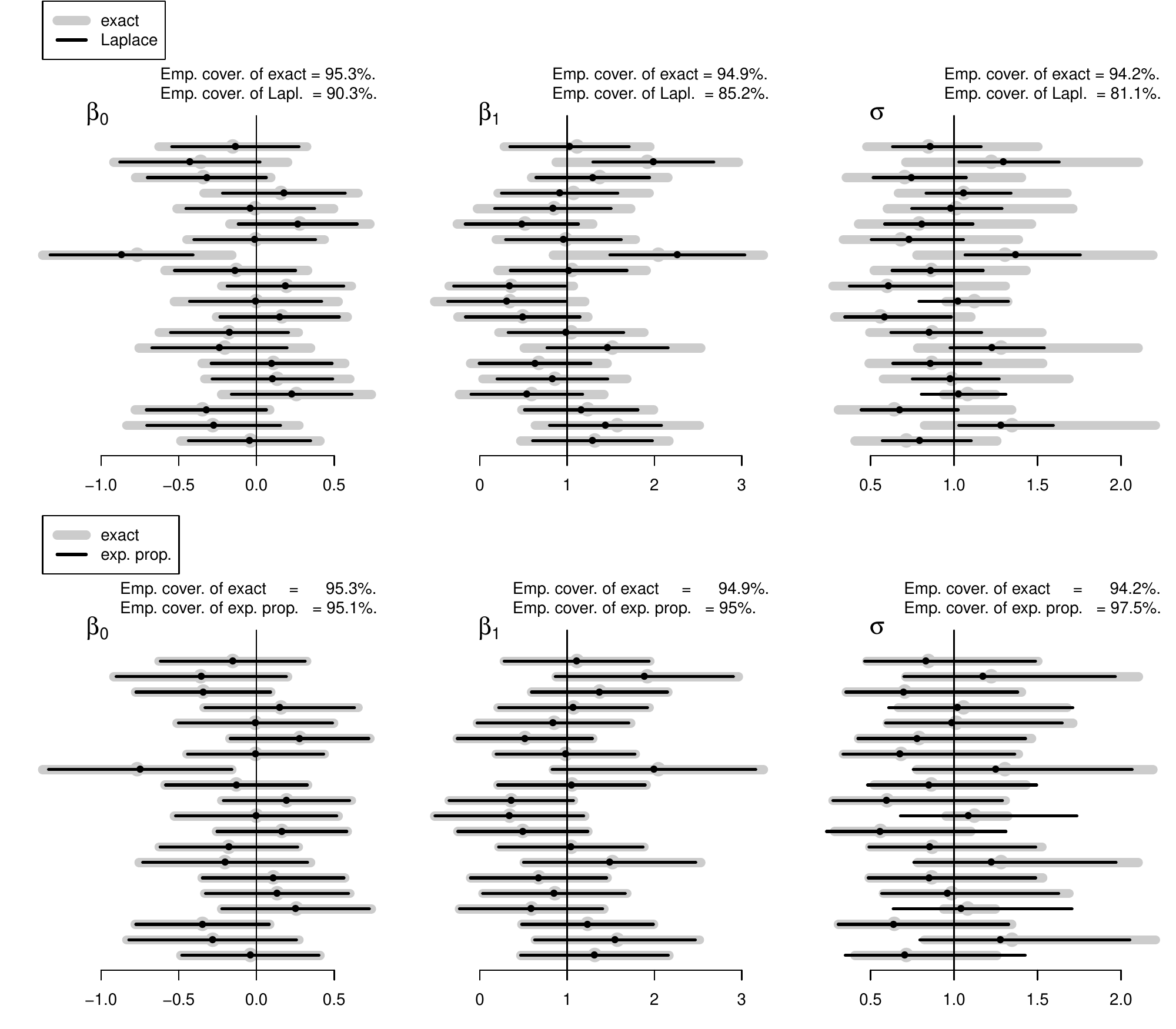}}
\caption{\it Comparison of point estimation and 95\% confidence 
interval coverage for the first simulation study with true parameter values
given by (\ref{eq:trueParmsFirst}). The upper row of panels compares
exact maximum likelihood with Laplace approximation.
The low row of panels compares
exact maximum likelihood with expectation propagation approximation.
The horizontal lines
indicate expectation propagation-based confidence intervals 
for 20 randomly chosen replications of the simulation study 
described in the text. The points indicate 
the corresponding approximate maximum likelihood estimates. 
The vertical lines indicate true parameter values. The percentages 
displayed at the top of each panel are empirical coverages over 
all $1,000$ replications for each method involved
in the comparison.
}
\label{fig:JASAsimStudy1CIplot} 
\end{figure}

For each simulated dataset, the probit mixed model defined by (\ref{eq:xFandxRsimOne})
was fit using each of the following approaches:
\begin{itemize}
\item[(1)] Exact maximum likelihood with adaptive Gauss-Hermite quadrature used 
for the univariate intractable integrals. This was achieved using the function
\texttt{glmer()} in the \textsf{R} package \texttt{lme4} (Bates \textit{et al.}, 2015).
The number of points for evaluation of the adaptive Gauss-Hermite approximation
was fixed at $100$.
\item[(2)] The Laplace approximation used by \texttt{glmer()}.
\item[(3)] Expectation propagation as described in Section \ref{sec:EPmain}.
\end{itemize}
Of interest is comparison of quadrature-free approximations 
(2) and (3) against the exact maximum likelihood benchmark.
Figure \ref{fig:JASAsimStudy1CIplot} contrasts the point estimates
and confidence intervals produced by Laplace approximation and expectation
propagation against those produced by exact maximum likelihood.
The first row of Figure \ref{fig:JASAsimStudy1CIplot}  
shows that Laplace approximation results in shoddy
statistical inference, with the empirical coverage values
falling well below the advertized 95\% level.
The gray line segments for exact likelihood confidence
intervals and black line segments for their
Laplace approximations have very noticeable discrepancies.
In the second row of Figure \ref{fig:JASAsimStudy1CIplot}
we repeat the empirical coverage percentages and gray
line segments for exact likelihood inference and, instead,
compare these results with those produced by expectation
propagation. For the fixed effects, $\beta_0$ and $\beta_1$, the
empirical coverage of expectation propagation
is seen to be very close to 95\%. For the standard deviation
parameter, $\sigma$, expectation propagation delivers slightly
more coverage than advertized (97.5\% versus 95\%). 
However, the relatively low sample sizes in this study should
be kept in mind. The simulation study in the next subsection
uses higher sample sizes and expectation propagation is seen
to be particularly accurate in terms of confidence interval 
coverage.

\subsubsection{Accuracy and Speed Assessment for Bivariate Random Effects}

In this study we simulated $1,000$ datasets according to 
a $\dR=2$ version of (\ref{eq:probitMixMod})
with true parameter values:
\begin{equation}
\bbetaTrue=[0.37,0.93,-0.46,0.08,-1.34,1.09]^T\quad\mbox{and}\quad
\bSigmaTrue=\left[
\begin{array}{rr}
0.53  & -0.36 \\
-0.36 & 0.92
\end{array}
\right].
\label{eq:trueParmsSecond}
\end{equation}
The number of groups was fixed at $m=250$ and each $n_i$ value
selected randomly from a discrete Uniform distribution on
$\{20,21,\ldots,30\}$. The $\bxF_{ij}$ and $\bxR_{ij}$ vectors
were of the form
$$\bxF_{ij}=[1,x_{1,ij},x_{2,ij},x_{3,ij},x_{4,ij},x_{5,ij}]^T\quad
\mbox{and}\quad
\bxR_{ij}=[1,x_{1,ij}]^T
$$
where each $x_{k,ij}$ was generated independently from a Uniform
distribution on the unit interval. All relative tolerance
values were set to $10^{-5}$ and the maximum number of iteration 
values were set to $100$, which is relevant for the upcoming
speed assessment.

\begin{figure}[!ht]
\centering
{\includegraphics[width=\textwidth]{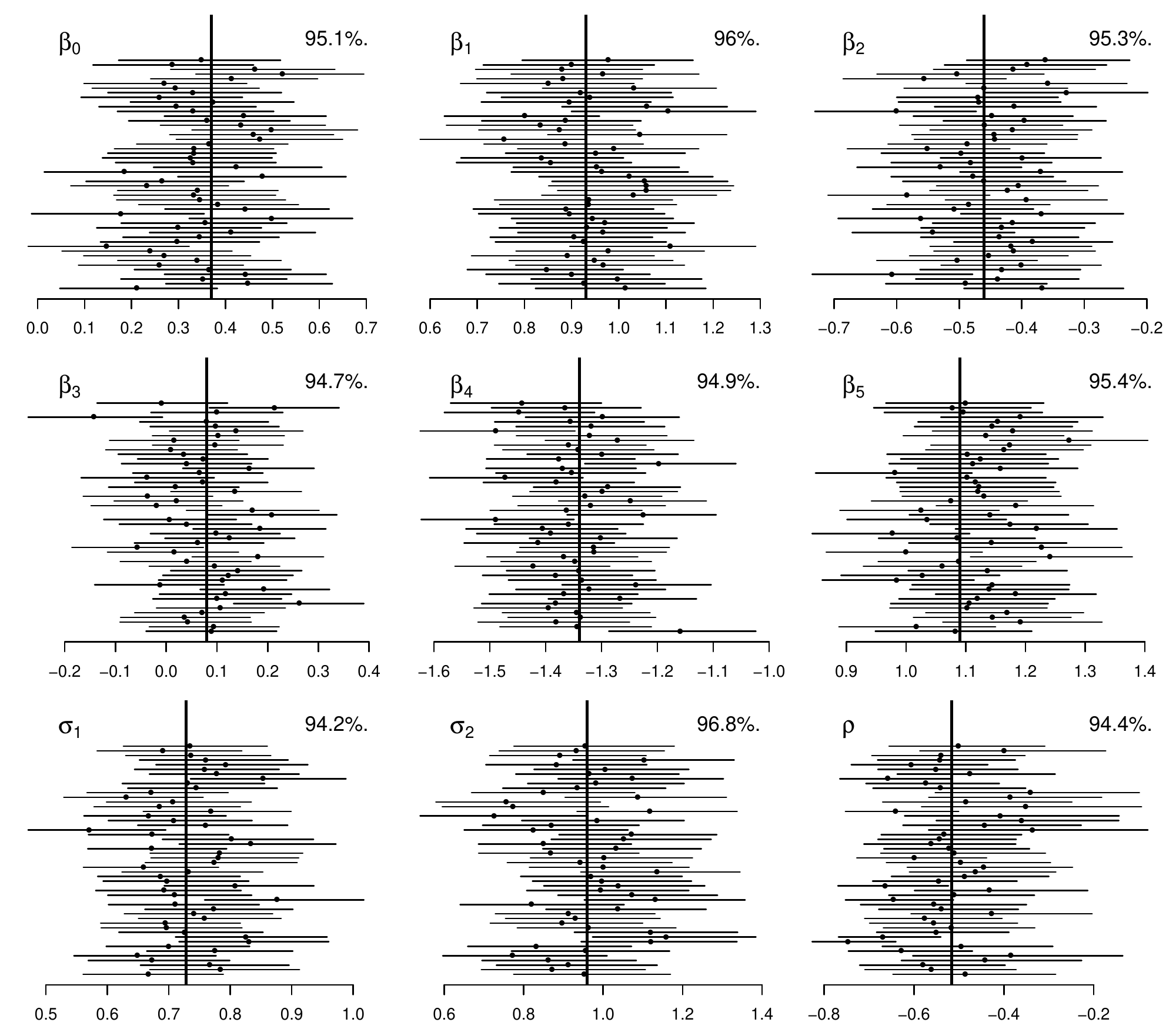}}
\caption{\it Summary of confidence interval coverage for the
second simulation study with true parameter values
given by (\ref{eq:trueParmsSecond}). The horizontal lines
indicate expectation propagation-based confidence intervals 
for 50 randomly chosen replications of the simulation study 
described in the text. The solid circular points indicate 
the corresponding point estimates. The vertical lines
indicate true parameter values. The percentage in the top right-hand
corner of each panel is the empirical coverage over all $1,000$ replications.
}
\label{fig:JASAsimStudy2CIplot} 
\end{figure}

The points and horizontal line segments in Figure \ref{fig:JASAsimStudy2CIplot} 
are displays of estimates and corresponding 95\% confidence intervals for each of 
the interpretable model parameters, for $50$ randomly chosen replications.
The numbers in the top right-hand corner of each panel are the empirical 
coverage values based on all $1,000$ replications.
For all nine parameters, the empirical coverage values are in keeping
with the advertized coverage of 95\%, and is an indication of
excellent accuracy for this setting.

Despite the higher samples and complexity of the model, we have gotten the
fitting times down to tens of seconds in the \textsf{glmmEP} package
within the \textsf{R} computing environment. This has been achieved by
implementation of Algorithm \ref{alg:mainAlgo} in a low level language
so that approximate likelihood evaluations are very rapid.
The computing speed depends upon various relative tolerance values and upper
bounds on numbers of iterations for the various iterative schemes as well as
attributes of the computer. This simulation study was run on a 
MacBook Air laptop with 8 gigabytes of random access memory
and a 2.2 gigahertz processor. The convergence stopping criteria
values are given earlier in this section.
Over the 1,000 replications the median computing time was 18 seconds,
the upper quartile was 20 seconds and the maximum was 34 seconds.
Such speed is impressive given that each data set contained tens of thousands
of observations and bivariate random effects are accurately handled.

\subsection{Application to Data from a Fertility Study}\label{sec:fertility}

Data from a 1988 Bangladesh fertility study are stored 
in the data frame \texttt{Contraception} within the 
\textsf{R} package \textsf{mlmRev} (Bates, Maechler and Bolker, 2014).
Steele, Diamond and Amin (1996) contains details of the study and 
some multilevel analyses. Variables in the \texttt{Contraception}
data frame include:
\begin{description}
\item[\texttt{use}] a two-level factor variable indicating whether 
a woman is a user of contraception at the time of the
survey, with levels \texttt{Y} for use and \texttt{N} for non-use,
\item[\texttt{age}] age of the woman in years at the time of the survey,
centred about the average age of all women in the study,
\item[\texttt{district}] a multi-level factor variable that codes
the district, out of 60 districts in total, in which the woman lives,
\item[\texttt{urban}] a two-level factor variable indicating whether
or not the district in which the woman lives is urban, with levels
\texttt{Y} for urban dwelling and \texttt{N} for rural dwelling, and
\item[\texttt{livch}] a four-level factor variable that indicates
the number of living children of the woman, with levels 
\texttt{0} for no children, \texttt{1} for one child, 
\texttt{2} for two children and \texttt{3+} for three or more
children.
\end{description}
A random intercepts and slopes probit mixed model for these data is
\begin{equation}
\setlength\arraycolsep{1pt}{
\begin{array}{rcl}
&&I(\texttt{use}_{ij}=\texttt{Y})|u_{0i},u_{1i}
\simind\mbox{Bernoulli}\Big(\Phi\big(\beta_0+u_{0i}+(\beta_1+u_{1i})
I(\texttt{urban}_{ij}=\texttt{Y})\\
&&\qquad +\beta_2\texttt{age}_{ij}
+\beta_3\,I(\texttt{livch}_{ij}=\texttt{1})+\beta_4\,I(\texttt{livch}_{ij}=\texttt{2})
+\beta_5\,I(\texttt{livch}_{ij}=\texttt{3+})\big)\Big)
\end{array}
}
\label{eq:contracModelOne}
\end{equation}
where $I(\Psc)=1$ if $\Psc$ is true and $0$ otherwise.
Also, $\texttt{use}_{ij}$ denotes the value of $\texttt{use}$ for the $j$th
woman within the $i$th district, $1\le i\le 60$, with the other variables
defined analogously.
The bivariate random effects vectors are assumed to satisfy
\begin{equation}
\left[
\begin{array}{c}
u_{0i}\\
u_{1i}
\end{array}
\right]\simind N\left(\left[\begin{array}{c}    
0\\
0
\end{array}
\right],
\left[\begin{array}{cc}    
\sigma_1^2           & \rho\sigma_1\sigma_2\\
\rho\sigma_1\sigma_2 & \sigma_2^2
\end{array}
\right]
\right).
\label{eq:contracModelTwo}
\end{equation}

We fitted this model using our expectation propagation approximate likelihood 
inference scheme. It took about 35 seconds on the fourth author's MacBook Air laptop 
(2.2 gigahertz processor and 8 gigabytes of random access memory) to 
produce the inferential summary given in Table \ref{tab:contracAnaRes}.

\begin{table}
\begin{center}
\begin{tabular}{crrr}
parameter   &95\% C.I. low.  &$\ \ \ \ \ \ \ \ \ $estimate   &  95\% C.I. upp.\\[0.2ex]
\hline\\[-1.6ex]
$\beta_0$   &$-1.2185$      &$-1.0418$         &$-0.8651$ \\
$\beta_1$   &$0.2956$       &$0.5003$         &$0.7049$  \\
$\beta_2$   &$-0.0259$      &$-0.0164$        &$-0.0068$ \\
$\beta_3$   &$0.4934$       &$0.6815$         &$0.8698$  \\
$\beta_4$   &$0.6223$       &$0.8306$         &$1.0389$  \\
$\beta_5$   &$0.6102$       &$0.8244$         &$1.0387$  \\
$\sigma_1$  &$0.2748$       &$0.3785$         &$0.5214$  \\
$\sigma_2$  &$0.3096$       &$0.4965$         &$0.7962$  \\
$\rho$      &$-0.9367$      &$-0.7984$        &$-0.4446$\\
\hline
\end{tabular}
\end{center}
\caption{\it Expectation propagation approximate maximum likelihood
estimates and corresponding 95\% confidence interval (C.I.) lower and
upper limits for the parameters in model (\ref{eq:contracModelOne})
and (\ref{eq:contracModelTwo}).}
\label{tab:contracAnaRes} 
\end{table}

Each of the parameters is seen to be statistically significantly different
from zero. As examples, the 95\% confidence interval for $\beta_2$ 
of $(0.296,0.705)$ indicates a higher use of contraception in
urban districts ad the 95\% confidence interval for $\sigma_2$ 
of $(0.310,0.796)$ shows that their is signficant heterogeneity 
in the urban versus rural effect across the 60 districts.

\begin{figure}[!ht]
\centering
{\includegraphics[width=0.6\textwidth]{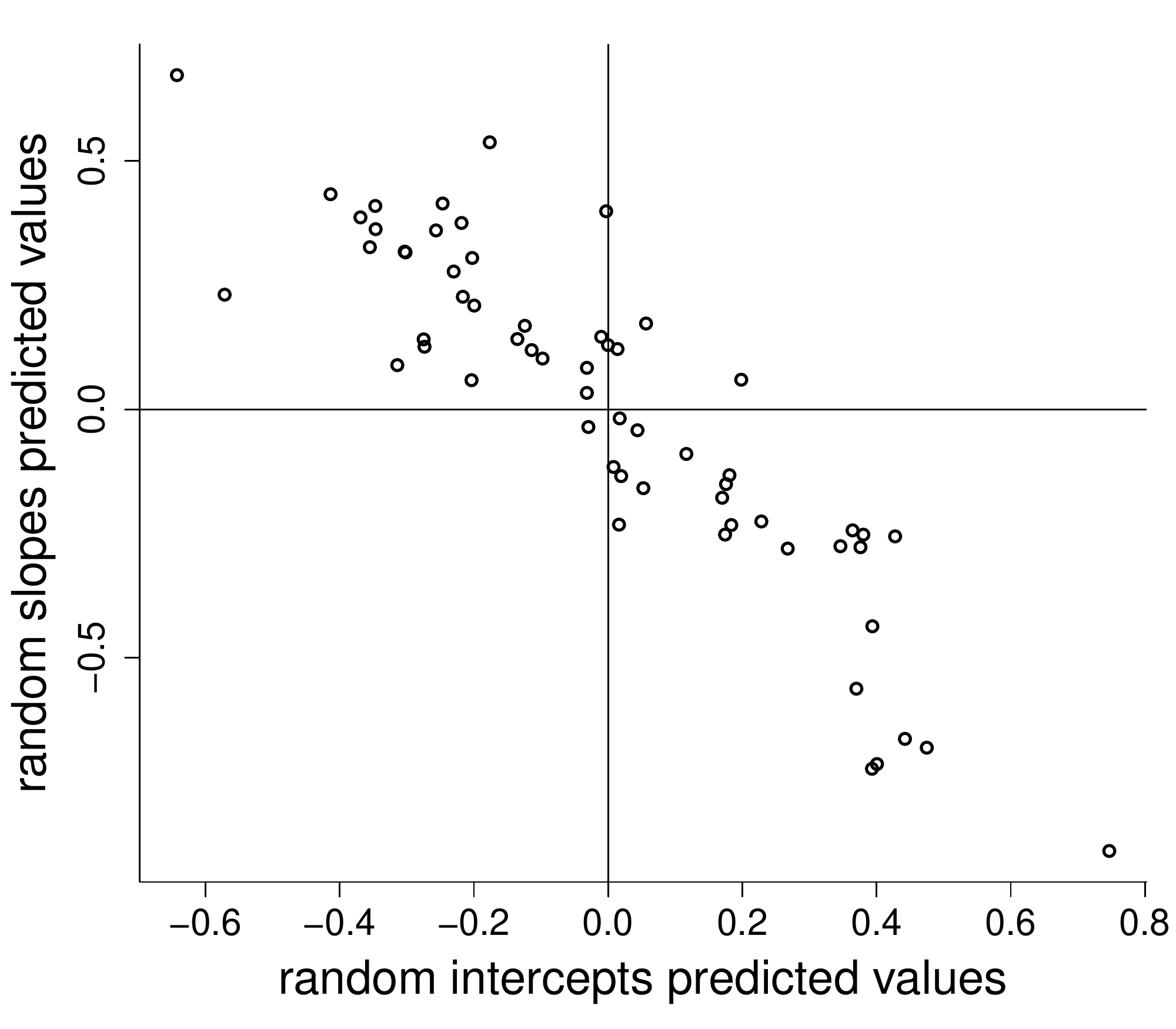}}
\caption{\it Scatterplot of the expectation propagation-approximate best predictions
of the random slopes and corresponding random intercepts for the fit of model given by
(\ref{eq:contracModelOne}) and (\ref{eq:contracModelTwo}) to data
from a 1998 Bangladesh fertility study.}
\label{fig:contracBestPreds} 
\end{figure}

We also used expectation propagation approximate best prediction
to obtain predictions of the $u_{0i}$ and $u_{1i}$ values.
The results are plotted in Figure \ref{fig:contracBestPreds} 
and provide a visualization of between-district heterogeneity.

\section{Theoretical Considerations}\label{sec:theorConsid}

We now discuss the question regarding whether the excellent inferential accuracy
of the Section \ref{sec:EPmain} methodology is supported by theory. A fuller theoretical
analysis is the subject of ongoing work involving the first four authors and,
upon completion, will be reported elsewhere. In this section we provide a heuristic
explanation for the accuracy of expectation propagation in the binary response
mixed model context.

First note that the $i$th log-likelihood summand is
$$\ell_i(\bbeta,\bSigma)=
\log\int_{\real^{d^R}}\left\{\prod_{j=1}^{n_i}\frac{p(\by_i|\bu_i;\bbeta)}
{\ptilde(\by_i|\bu_i;\bbeta) }\right\}
\exp\left\{
\left[
\begin{array}{c}        
1\\
\bu_i\\
\vech(\bu_i\bu_i^T)
\end{array}
\right]^T{\widehat\bdetaEP}_i
\right\}\,d\bu_i
$$
where $\ptilde(\by_i|\bu_i;\bbeta)$ is given by expression (\ref{eq:pEPexpn})
with the $\bdeta_{ij}$ set to the converged $\etaSUBpyijTOui$
values. We also have
$$\ellEP_i(\bbeta,\bSigma)=
\log\int_{\real^{d^R}}
\exp\left\{
\left[
\begin{array}{c}        
1\\
\bu_i\\
\vech(\bu_i\bu_i^T)
\end{array}
\right]^T{\widehat\bdetaEP}_i
\right\}\,d\bu_i.
$$
Now make the change of variables 
$$\bv=\bDelta_i^{-1}\{\bu_i - \BPEP(\bu_i)\}\quad\mbox{where}\quad 
\bDelta_i\equiv\CovEP(\bu_i|\by)^{1/2}
$$
involving the expectation propagation-approximate 
best predictor quantities given by (\ref{eq:BPdefn}) 
and (\ref{eq:sampVersion}).
Straightforward manipulations then lead to the discrepancy 
between $\ell(\bbeta,\bSigma)$ and $\ellEP_i(\bbeta,\bSigma)$
equalling
\begin{equation}
\ell(\bbeta,\bSigma)-\ellEP_i(\bbeta,\bSigma)=\log\int_{\real^{d^R}}
\left\{\prod_{j=1}^{n_i}A_{ij}(\bDelta_i\bv)\right\}\phi_{\bI}(\bv)\,d\bv
\label{eq:discrepOne}
\end{equation}
where, for any $\bx\in\real^{\dR}$, 
$\phi_I(\bx)\equiv(2\pi)^{-\dR/2}\exp(-\smhalf\bx^T\bx)$ and
\begin{eqnarray*}
&&A_{ij}(\bx)\equiv F\Big((2y_{ij}-1)\big(\bbeta^T\bxF_{ij}+(\BPEP(\bu_i)+\bx)^T\bxR_{ij}\big)\Big)\\[1ex]
&&\qquad\qquad\qquad\times\exp\left\{
-\left[
\begin{array}{c}        
1\\
\BPEP(\bu_i)+\bx\\
\vech\Big(\big(\BPEP(\bu_i)+\bx\big)\big(\BPEP(\bu_i)+\bx\big)^T\Big)
\end{array}
\right]^T      
\etaSUBpyijTOui\right\}.
\end{eqnarray*}
Using the same change of variables, the moment-matching conditions corresponding
to the Kullback-Leibler projection (\ref{eq:KLijth}) are
\begin{equation}
\int_{\real^{\dR}}\bv^{\otimes\,k}A_{ij}(\bDelta_i\bv)\phi_I(\bv)\,d\bv=
\int_{\real^{\dR}}\bv^{\otimes\,k}\phi_I(\bv)\,d\bv,\quad k=0,1,2,
\label{eq:momConds}
\end{equation}
where $\bv^{\otimes\,0}\equiv1$, $\bv^{\otimes\,1}\equiv\bv$ and
$\bv^{\otimes\,2}=\bv\bv^T$.

To aid intuition, for the remainder of this section we restrict attention to 
$\dR=1$ and write $\delta_i$ instead of $\bDelta_i$ to signify the
fact that this quantity is scalar in this special case. 
Next, we make the 
\begin{equation}
\mbox{working assumption:}\quad \delta_i=O_p(n_i^{-1/2}).
\label{eq:workAss}
\end{equation}
This assumption is in keeping with the fact that $\delta_i$ is the expectation propagation 
approximation to the sample standard deviation of $\BPEP(\bu_i)-\bu_i$. Then Taylor series
expansion of $A_{ij}$ about zero and substitution into the $\dR=1$ version of (\ref{eq:momConds})
leads to 
$$A_{ij}(0)=1+O(\delta_i^4),\quad A'_{ij}(0)=O(\delta_i^2)\quad\mbox{and}\quad
A''_{ij}(0)=O(\delta_i^2).
$$
Plugging these into (\ref{eq:discrepOne}) and using $\log(1+\varepsilon)\approx\varepsilon$
for small $\varepsilon$ we obtain 
$$\ell_i(\bbeta,\bSigma)-\ellEP_i(\bbeta,\bSigma)=O_p(n_i^{-1/2})\quad\mbox{under}\ (\ref{eq:workAss}).$$

These heuristics suggest that expectation propagation 
provides consistent estimation of the log-likelihood summands 
as the number of measurements in the $i$th group increases.
The deeper question concerning the asymptotic statistical properties
of the expectation propagation-based estimators 
(${\widehat\bbetaEP},{\widehat\bSigmaEP})$ requires 
more delicate theoretical analysis. As mentioned earlier in this section,
this question is being pursued by authors of this article.

Before closing this section, we mention that there is a small but
emerging body of research concerning the large sample behavior
of expectation propagation for approximation Bayesian inference.
A recent contribution of this type is Dehaene \myand Barthelm\'e (2018)
which provides Bernstein-von Mises theory for Bayesian expectation
propagation.

\section{Higher Level and Crossed Random Effects Extensions}\label{sec:highLevCrossed}

The binary mixed model given by (\ref{eq:probitMixMod}) is adequate for the
common situation of there being only one grouping mechanism. However, more
elaborate models are required for situations such as hierarchical and
cross-tabulated grouping mechanisms. Goldstein (2010), for example, 
provides an extensive treatment of mixed models with higher levels of
nesting. A major reference for crossed random effects mixed models is
Baayen, Davidson \myand Bates (2008). Here we provide advice regarding
extension our expectation propagation approach to these settings.

\begin{figure}[!ht]
\centering
{\includegraphics[width=0.65\textwidth]{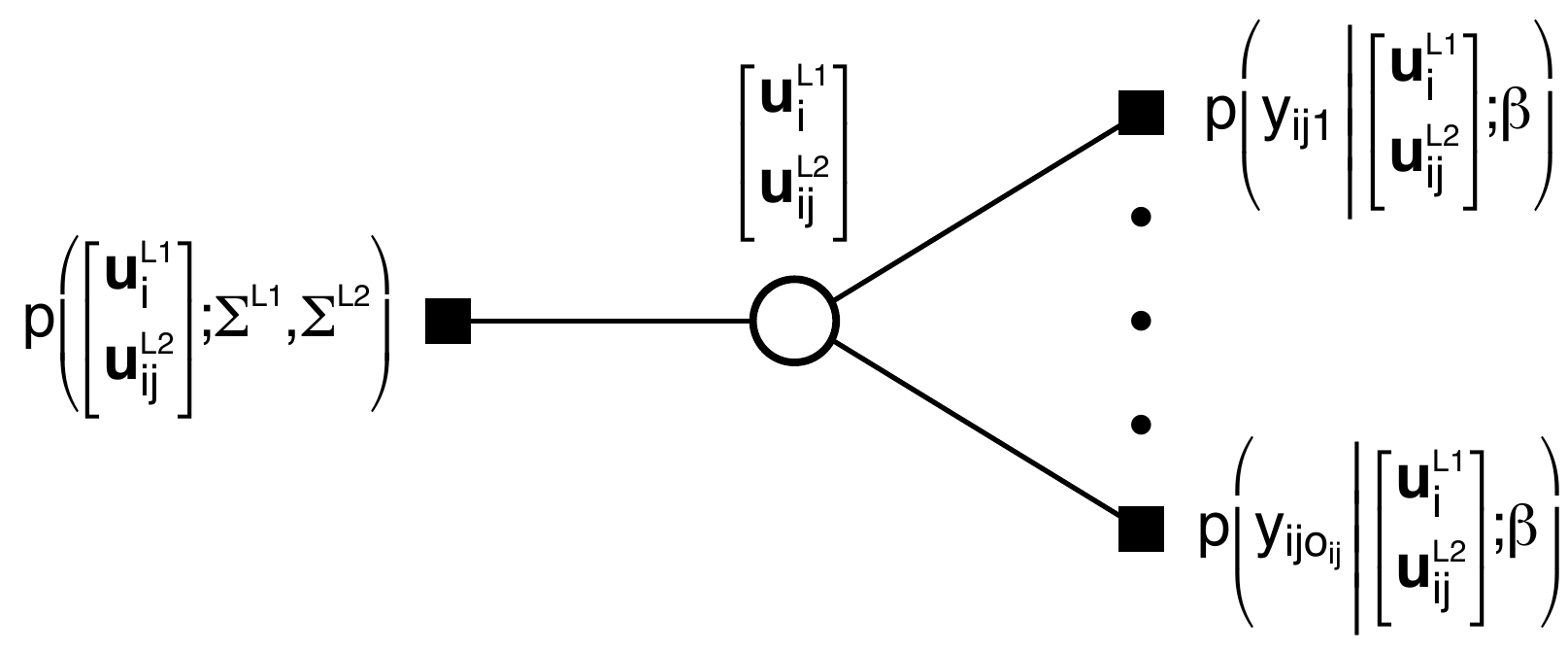}}
\caption{\it Factor graph representation of the product structure of
the integrand in (\ref{eq:elli}). The open circle corresponds to the 
random effect vector $[\buLone_i\ \buLtwo_{ij}]^T$ and the solid 
rectangles indicate factors in the integrand of 
(\ref{eq:twoLevelLogLik}).
Edges indicate dependence of each factor on 
$[\buLone_i\ \buLtwo_{ij}]^T$.}
\label{fig:twoLevFacGraph} 
\end{figure}

The \emph{two levels of nesting} extension of (\ref{eq:probitMixMod}) is
\begin{equation}
\begin{array}{c}
y_{ijk}|\buLone_i,\buLtwo_{ij}\simind\mbox{Bernoulli}\Big(F\big(\bbeta^T\bxF_{ijk}
+(\buLone_i)^T\bxRone_{ijk}
+(\buLtwo_{ij})^T\bxRtwo_{ijk}\big)\Big),\\[2ex]
\buLone_i\simind N(\bzero,\bSigmaLone)\ \mbox{independently of}\ 
\buLtwo_{ij}\simind N(\bzero,\bSigmaLtwo),\\[2ex]
1\le i\le m,\quad 1\le j\le n_i,\quad 1\le k\le o_{ij}.
\end{array}
\label{eq:probitMixModHigher}
\end{equation}
The response $y_{ijk}$ and predictor vectors $\bxF_{ijk}$, $\bxRone_{ijk}$
and $\bxRtwo_{ijk}$ correspond to the $k$th set of measurements within the
$j$th inner group within the $i$th outer group. The number of outer groups
is $m$ and the number of inner groups in the $i$th outer group is $n_i$.
The sample size of the $j$th group in the $i$th outer group is $o_{ij}$.
Also, $\bxRone_{ijk}$ is $\dRone\times1$ and $\bxRtwo_{ijk}$ is $\dRtwo\times1$.
The log-likelihood of $(\bbeta,\bSigmaLone,\bSigmaLtwo)$ may be written as
\begin{equation}
\ell(\bbeta,\bSigmaLone,\bSigmaLtwo)
=\sumim\log\int_{\real^{\dR}}\prod_{j=1}^{n_i}\prod_{k=1}^{o_{ij}}
p\left(y_{ijk}\Bigg|\buLoneLtwoij;\bbeta\right)
p\left(\buLoneLtwoij;\bSigmaLone,\bSigmaLtwo\right)
\,d\buLoneLtwoij
\label{eq:twoLevelLogLik}
\end{equation}
where $\dR=\dRone+\dRtwo$,
$$p\left(y_{ijk}\Bigg|\buLoneLtwoij;\bbeta\right)\equiv
F\Big((2y_{ijk}-1)\big(\bbeta^T\bxF_{ijk}
+(\buLone_i)^T\bxRone_{ijk}+(\buLtwo_{ij})^T\bxRtwo_{ijk}\big)\Big),\ y_{ijk}=0,1,
$$
and 
$$p\left(\buLoneLtwoij;\bSigmaLone,\bSigmaLtwo\right)\equiv
|2\pi\bSigmaLone|^{-1/2}|2\pi\bSigmaLtwo|^{-1/2}
\exp\left\{-\smhalf\buLoneLtwoij^T
\left[\begin{array}{cc}          
\bSigmaLone & \bzero \\
\bzero      & \bSigmaLtwo
\end{array}\right]^{-1}\buLoneLtwoij\right\}.
$$
Expectation propagation approximation of $\ell(\bbeta,\bSigmaLone,\bSigmaLtwo)$ then
proceeds by message passing on the factor graph displayed in Figure \ref{fig:twoLevFacGraph}.
In the probit case Theorem 1 can be called upon to obtain closed form updates for
the message natural parameter vectors leading to an algorithm analogous to 
Algorithm \ref{alg:mainAlgo}.

A \emph{crossed random effects} extension of (\ref{eq:probitMixMod}) is
$$
\begin{array}{c}
y_{ii'j}|\buC_i,\buCdash_{i'}\simind\mbox{Bernoulli}\Big(F\big(\bbeta^T\bxF_{ii'j}
+(\buC_i)^T\bxR_{ii'j}
+(\buCdash_{i'})^T\bxRdash_{ii'j}\big)\Big),\\[2ex]
\buC_i\simind N(\bzero,\bSigma)\ \mbox{independently of}\ 
\buCdash_{i'}\simind N(\bzero,\bSigma'),\\[2ex]
1\le i\le m,\quad 1\le i'\le m',\quad 1\le j\le n_{ii'}
\end{array}
$$
where the data are cross-tabulated according to membership
of two groups of sizes $m$ and $m'$ indexed according to the pair 
$(i,i')\in\{1,\ldots,m\}\times\{1,\ldots,m'\}$,
with $n_{ii'}$ denoting the sample size within group $(i,i')$.
Note that $n_{ii'}=0$ is a possibility for some $(i,i')$.
The response $y_{ii'j}$ and the predictor vectors $\bxF_{ii'j}$, 
$\bxR_{ii'j}$ and $\bxRdash_{ii'j}$ correspond to the $j$th
set of measurements within group $(i,i')$.
The $\bu_i$, $1\le i\le m$, are $\dR\times 1$ random effects
for group-specific departures from the fixed effects for
the first group.
The $\bu'_i$, $1\le i\le m'$, are $\dRdash\times 1$ random effects
for group-specific departures from the fixed effects for
the second group. The log-likelihood of $(\bbeta,\bSigma,\bSigma')$ is
$$
\ell(\bbeta,\bSigma,\bSigma')
=\sumim\sum_{i'=1}^{m'}I(n_{ii'}>0)\log\int_{\real^{\dR+\dRdash}}\prod_{j=1}^{n_{ii'}}
p\left(y_{ii'j}\Bigg|\bubuDashiidj;\bbeta\right)
p\left(\bubuDashiidj;\bSigma,\bSigma'\right)
\,d\bubuDashiidj
$$
where 
$$p\left(y_{ii'j}\Bigg|\bubuDashiidj;\bbeta\right)\equiv
F\Big((2y_{ijk}-1)\big(\bbeta^T\bxF_{ii'j}
+(\buC_i)^T\bxR_{ii'j}
+(\buCdash_{i'})^T\bxRdash_{ii'j}\big)\Big),\ y_{ijk}=0,1,
$$
and
$$p\left(\bubuDashiidj;\bSigma,\bSigma'\right)\equiv
|2\pi\bSigma|^{-1/2}|2\pi\bSigma'|^{-1/2}
\exp\left\{-\smhalf\bubuDashiidj^T
\left[\begin{array}{cc}          
\bSigma & \bzero \\
\bzero      & \bSigma'
\end{array}\right]^{-1}\bubuDashiidj\right\}.
$$
Expectation propagation approximation of $\ell(\bbeta,\bSigma,\bSigma')$
can be achieved via message passing on a factor graph similar to that shown in
Figure \ref{fig:twoLevFacGraph}.

\section{Transferral to Other Mixed Models}\label{sec:otherGLMMs}

Until now we have mainly focused on the special case of probit
mixed models with Gaussian random effects since the requisite 
Kullback-Leibler projections have closed form solutions. However, 
our approach is quite general and, at least in theory, applies to other mixed
models. We now briefly describe transferral to other mixed models.

\subsection{Logistic Mixed Models}

As we mention in Section \ref{sec:binMixMod}, the probit and logistic
cases are distinguished according to whether $F=\Phi$ or $F=\expit$.
Therefore, transferral from probit to logistic mixed models involves
replacement of $\fInput$ in Theorem 1 by 
\begin{equation}
\fInput(\bx)=\expit(c_0+\bc_1^T\bx)\exp\left\{  
\left[\begin{array}{c}
\bx\\
\vech(\bx\bx^T)
\end{array}
\right]^T
\left[  
\begin{array}{c}
\bdetaOneInput\\[1ex]
\bdetaTwoInput
\end{array}
\right]
\right\},\quad \bx\in\real^d.
\label{eq:expitfInput}
\end{equation}
In view of Lemma 1 of the online supplement, 
Kullback-Leibler projection of $\fInput$ onto the unnormalized Normal
family involves univariate integrals of the form
\begin{equation}
\infint\,x^p\exp\{qx - rx^2-\log(1+e^x)\}\,dx,\quad p=0,1,2,\ q\in\real,\ r>0.
\label{eq:scrCfam}
\end{equation}
In the Bayesian context, Gelman \textit{et al.} (2014; Section 13.8) and Kim \myand Wand (2017)
describe quadrature-based approaches to evaluation of (\ref{eq:scrCfam}),
each of which transfers to the frequentist context dealt with here.
However, there is a significant speed cost compared with the probit case.

An alternative approach involves use of the family of approximations to 
$\expit$ of the form 
$$\expit_k(x)\equiv\sum_{i=1}^k\,p_{k,i}\Phi(s_{k,i}x)$$
for constants $p_{k,i}$ and $s_{k,i}$, as advocated by Monahan \myand Stefanski (1989).
Since the approximation is a linear combination of scalings of $\Phi$,
the function $\fInput$ in (\ref{eq:expitfInput}) with $\expit$ replaced by $\expit_k$
admits closed form Kullback-Leibler projections onto the unnormalized
Multivariate Normal, leading to fast and accurate inference for logistic
mixed models. 

Details on the mechanics and performance of expectation propagation
for logistic mixed models is to be reported in Yu (2019). 

\subsection{Other Generalized Linear Mixed Models}

Whilst we have focused on the binary response situation in this article,
we quickly point out that the principles apply to other generalized 
linear mixed models such as those based on the Gamma and Poisson 
families. Note that (\ref{eq:logLikFirst}) with $F=\expit$ generalizes
to
\begin{eqnarray*}
&&\ell(\bbeta,\bSigma)=\sum_{i=1}^m\log
\int_{\dR}\left[\prod_{j=1}^{n_i}\exp\Big\{y_{ij}(\bbeta^T\bxF_{ij}+\bu^T\bxR_{ij})
-b\big(\bbeta^T\bxF_{ij}+\bu^T\bxR_{ij}\big)+c(y_{ij})\Big\}\right]\\[1ex]
&&\qquad\qquad\times|2\pi\bSigma|^{-1/2}\exp(-\smhalf\bu^T\bSigma^{-1}\bu)\,d\bu
\end{eqnarray*}
where the functions $b$ and $c$ are as given in Table 2.1 of
McCullagh \myand Nelder (1989). Setting $b(x)=\log(1+e^x)$ and $c(x)=0$
gives the $F=\expit$ logistic mixed model while putting $b(x)=e^x$ and $c(x)=-\log(x!)$
gives the corresponding Poisson mixed model. The family of integrals
$$
\infint\,x^p\exp\{qx - rx^2-b(x)\}\,dx,\quad p=0,1,2,\ q\in\real,\ r>0,
$$
is required to facilitate the required Kullback-Leibler projections for the $\dR=1$
case. With the exception of logistic mixed models, multivariate numerical integration 
appears to be required when $\dR>1$. Yu (2019) will contain a detailed account of the 
practicalities and performance of expectation propagation for this class of models.

\section*{Acknowledgments}

This research was supported by Australian Research Council
Discovery Project DP180100597. We are grateful for advice from 
Jim Booth, Omar Ghattas and Alan Huang on aspects of this research.

\section*{References}

\bib
Azzalini, A. (2017). The \textsf{R} package \texttt{sn}: The skew-normal and skew-t
 distributions (version 1.5). \texttt{http://azzalini.stat.unipd.it/SN}

\bib
Baayen, R.H., Davidson, D.J. and Bates, D.M. (2008).
Mixed-effects modeling with crossed random effects for
subjects and items. \textit{Journal of Memory and Language},
{\bf 59}, 390--412.

\bib
Bates, D., Maechler, M. and Bolker, B. (2014).
\textsf{mlmRev}: Examples from multilevel modelling software review. 
\textsf{R} package version 1.0.\\
\texttt{http://cran.r-project.org}.

\bib
Baltagi, B.H. (2013). 
\textit{Econometric Analysis of Panel Data, Fifth Edition}.
Chichester, U.K.: John Wiley \& Sons.

\bib
Bates, D., Maechler, M., Bolker, B. and Walker, S. (2015).
Fitting linear mixed-effects models using \textsf{lme4}. 
\textit{Journal of Statistical Software}, \textbf{67(1)}, 
1--48.

\bib
Bishop, C.M. (2006). \textit{Pattern Recognition and Machine Learning.}
New York: Springer.

\bib
Broyden, C.G. (1970). The convergence of a class of double-rank minimization
algorithms. \textit{Journal of the Institute of Mathematics and Its Applications},
\textbf{6}, 76--90.

\bib
Carlin, B.P. and Gelfand, A.E. (1991). A sample reuse method for accurate parametric
empirical Bayes confidence intervals. \textit{Journal of the Royal Statistical 
Society, Series B}, {\bf 53}, 189--200.

\bib
Dehaene, G. and Barthelm\'e, S. (2018). Expectation propagation in the 
large-data limit. \textit{Journal of the Royal Statistical Society, Series B},
{\bf 80}, 199--217.

\bib
Diggle, P., Heagerty, P., Liang, K.-L. and Zeger, S. (2002).
{\it Analysis of Longitudinal Data, Second Edition}.
Oxford, U.K.: Oxford University Press.

\bib
Fletcher, R. (1970). A new approach to variable metric algorithms.
\textit{Computer Journal}, \textbf{13}, 317--322.

\bib
Gelman, A., Carlin, J.B., Stern, H.S., 
Dunson, D.B.,Vehtari, A. and Rubin, D.B. (2014). 
\textit{Bayesian Data Analysis, Third Edition}, Boca Raton, Florida: CRC Press.

\bib
Gelman, A. and Hill, J. (2007).
\textit{Data Analysis using Regression and Multilevel/Hierarchical
  Models}.
New York: Cambridge University Press. 

\bib
Givens, G.H. and Hoetig, J.A. (2005).
\textit{Computational Statistics}, Hoboken, New Jersey: John Wiley \& Sons.

\bib
Goldfarb, D. (1970). A family of variable metric updates derived by 
variational means. \textit{Mathematics of Computation}, {\bf 24}, 
23--26.

\bib
Goldstein, H. (2010). \textit{Multilevel Statistical Models, Fourth 
Edition}. Chichester, U.K.: John Wiley \& Sons. 

\bib
Harville, D.A. (2008). \textit{Matrix Algebra from a Statistician's
Perspective}. New York: Springer.

\bib
Kim, A.S.I. and Wand, M.P. (2017).
On expectation propagation for generalised, linear and mixed models. 
\textit{Australian and New Zealand Journal of Statistics},
{\bf 59}, in press.

\bib
Magnus, J.R. and  Neudecker, H. (1999). \textit{Matrix Differential
Calculus with Applications in Statistics and Econometrics, Revised Edition}.
Chichester U.K.: Wiley.

\bib
Mardia, K.V., Kent, J.T. and Bibby, J.M. (1979). \textit{Multivariate Analysis}.
London: Academic Press.

\bib
The Mathworks Incorporated (2018). Natick, Massachusetts, U.S.A.

\bib
McCullagh, P. and Nelder, J.A. (1989). {\it Generalized Linear
Models, Second Edition}. London: Chapman and Hall.

\bib
McCulloch, C.E., Searle, S.R. and Neuhaus, J.M. (2008). \textit{Generalized, Linear,
and Mixed Models, Second Edition}. New York: John Wiley \& Sons.

\bib
Minka, T.P. (2001).
Expectation propagation for approximate Bayesian inference.
In J.S. Breese \myand D. Koller (eds),
\textit{Proceedings of the Seventeenth Conference on Uncertainty in
Artificial Intelligence}, pp. 362--369.
Burlington, Massachusetts: Morgan Kaufmann.

\bib
Minka, T. (2005). 
Divergence measures and message passing.
\textit{Microsoft Research Technical Report Series}, 
{\bf MSR-TR-2005-173}, 1--17.

\bib
Minka, T. \myand  Winn, J. (2008), 
Gates: A graphical notation for mixture models.
\textit{Microsoft Research Technical Report Series}, 
{\bf MSR-TR-2008-185}, 1--16.

\bib
Monahan, J.F. \myand Stefanski, L.A. (1989).
Normal scale mixture approximations to $F^*(z)$
and computation of the logistic-normal integral.
In Balakrishnan, N. (editor),
\textit{Handbook of the Logistic Distribution}.
New York: Marcel Dekker, 529--540.

\bib
Nelder, J.A. and Mead, R. (1965).
A simplex method for function minimization.
\textit{Computer Journal}, {\bf 7}, 308--313.

\bib
Pinheiro, J.C. and Bates, D.M. (2000).
{\it Mixed-Effects Models in S and S-PLUS}.
New York: Springer.

\bib
\textsf{R} Core Team (2018). \textsf{R}: A language and environment
for statistical computing. \textsf{R} Foundation for
Statistical Computing, Vienna, Austria.
\texttt{https://www.R-project.org/}.

\bib
Rao, J.N.K. and Molina, I. (2015).
\textit{Small Area Estimation, Second Edition}.
Hoboken, New Jersey: John Wiley \& Sons.

\bib
Rue, H., Martino, S. and Chopin, N. (2009).
Approximate Bayesian inference for latent Gaussian
models by using integrated nested Laplace approximations
(with discussion).
\textit{Journal of the Royal Statistical Society, Series B},
{\bf 71}, 319--392.

\bib
Shanno, D.F. (1970). Conditioning of quasi-Newton methods for function minimization.
\textit{Mathematics of Computation}, \textbf{24}, 647--656.

\bib
Steele, F., Diamond, I. and Amin, S. (1996). 
Immunization uptake in rural Bangladesh: a multilevel
analysis. \textit{Journal of the Royal Statistical Society,
Series A}, \textbf{159}, 289--299.

\bib
The Mathworks Incorporated (2018). Natick, Massachusetts, U.S.A.

\bib
Wainwright, M.J. and Jordan, M.I. (2008).
Graphical models, exponential families, and variational inference.
\textit{Foundations and Trends in Machine Learning}, {\bf 1}, 1--305.

\bib
Wand, M.P. and Ormerod, J.T. (2012).
Continued fraction enhancement of Bayesian computing.
\textit{Stat}, {\bf 1}, 31--41.

\bib
Wand, M.P. and Yu, J.C.F. (2018). \textsf{glmmEP}:
Fast and accurate likelihood-based inference in generalized
linear mixed models via expectation propagation.
\textsf{R} package version 0.999. \texttt{http://cran.r-project.org}.

\bib
Yu, J.C.F. (2019). \textit{Fast and Accurate Frequentist Generalized Linear
Mixed Model Analysis via Expectation Propagation.} Doctor of Philosophy
thesis, University of Technology Sydney.

\vfill\eject
%
%
\renewcommand{\theequation}{S.\arabic{equation}}
\renewcommand{\thesection}{S.\arabic{section}}
\renewcommand{\thetable}{S.\arabic{table}}
\setcounter{equation}{0}
\setcounter{table}{0}
\setcounter{section}{0}
\setcounter{page}{1}
\setcounter{footnote}{0}

\begin{center}

{\Large Supplement for:}
\vskip3mm

\centerline{\Large\bf Fast and Accurate Binary Response Mixed Model}
\vskip2mm
\centerline{\Large\bf Analysis via Expectation Propagation}
\vskip7mm
\centerline{\normalsize\sc By P. Hall$\null^1$, I.M. Johnstone$\null^2$, 
J.T. Ormerod$\null^3$, M.P. Wand$\null^4$ and J.C.F. Yu$\null^4$}
\vskip5mm
\centerline{\textit{$\null^1$University of Melbourne, 
$\null^2$Stanford University, $\null^3$University of Sydney}}
\vskip1mm
\centerline{\textit{and $\null^4$University of Technology Sydney}}
\vskip6mm

\end{center}

\section{Proof of Theorem 1}\label{sec:proofThmOne}

For $\bx\in\real^d$, define 
$$\phi_{\bSigma}(\bx)\equiv(2\pi)^{-d/2}|\bSigma|^{-1/2}\exp\big(-\smhalf\bx^T\bSigma^{-1}\bx\big)$$
so that
$$\phi_{\bI}(\bx)=(2\pi)^{-d/2}\exp\big(-\smhalf\bx^T\bx\big).$$
We continue to use an unadorned $\phi$ to denote the Univariate Normal
density function:
$$\phi(x)=(2\pi)^{-1/2}\exp(-\smhalf\,x^2).$$
The notation $\Vert\bv\Vert=\sqrt{\bv^T\bv}$ for 
a column vector $\bv$ is also used.

\jump
{\bf Lemma 1.} \textit{For any function $g:\real\to\real$ and $d\times1$ vectors
$\balpha_1$, $\balpha_2$ and $\balpha_3$ such that the integrals exist:}

\setlength\arraycolsep{1pt}{
\begin{eqnarray}
\int_{\real^d}g(\balpha_1^T\bx)\phi_{\bI}(\bx)\,d\bx&=&\infint g(\Vert\balpha_1\Vert\,z)\phi(z)\,dz,\\[2ex]
\int_{\real^d}g(\balpha_1^T\bx)(\balpha_2^T\bx)\phi_{\bI}(\bx)\,d\bx&=&
\{(\balpha_1^T\balpha_2)/\Vert\balpha_1\Vert\}
\infint z\,g(\Vert\balpha_1\Vert\,z)\phi(z)\,dz\\[2ex]
\textit{and}\ \int_{\real^d}g(\balpha_1^T\bx)(\balpha_2^T\bx)(\balpha_3^T\bx)\phi_{\bI}(\bx)\,d\bx&=&
(\balpha_2^T\balpha_3)\infint g(\Vert\balpha_1\Vert\,z)\phi(z)\,dz\\\label{eq:LemmaOneC}
&&\hskip-30mm+\{(\balpha_1^T\balpha_2)(\balpha_1^T\balpha_3)/\Vert\balpha_1\Vert^2\}
\infint (z^2-1)g(\Vert\balpha_1\Vert\,z)\phi(z)\,dz.\nonumber
\end{eqnarray}
}

\jump
{\bf Proof of Lemma 1.} Lemma 1 is a consequence of the fact that the integrals
on the left-hand side are, respectively,
$$E\{g(\balpha_1^T\bx)\},\ E\{g(\balpha_1^T\bx)(\balpha_2^T\bx)\}\quad\mbox{and}\quad
E\{g(\balpha_1^T\bx)(\balpha_2^T\bx)(\balpha_3^T\bx)\}
$$
where
$$\bx\sim\mbox N(\bzero_d,\bI_d).$$
We now focus on simplification of the third integral (\ref{eq:LemmaOneC}). 
Simplication of the first and second integrals is similar and simpler.
Make the change of variables
$$\bs\equiv\left[
\begin{array}{c}
s_1\\
s_2\\
s_3
\end{array}
\right]
=\bA\bx
\quad\mbox{where}\quad
\bA\equiv\left[
\begin{array}{c}
\balpha_1^T\\
\balpha_2^T\\
\balpha_3^T
\end{array}
\right]
$$
so that
$$E\{g(\balpha_1^T\bx)(\balpha_2^T\bx)(\balpha_3^T\bx)\}=E\{g(s_1)s_2s_3\}
\quad\mbox{where}\quad \bs\sim N(\bzero_3,\bA\bA^T).$$
We then note that,
\setlength\arraycolsep{1pt}{
\begin{eqnarray*}
E\{g(s_1)s_2s_3\}&=&\infint g(s_1)\left\{\infint\infint s_2s_3p(s_2s_3|s_1)ds_2ds_3\right\} p(s_1)ds_1\\
                 &=&\infint g(s_1)\left\{\infint\infint
\{\Cov(s_2,s_3|s_1)+E(s_2|s_1)E(s_3|s_1)\}ds_2ds_3\right\} p(s_1)ds_1
\end{eqnarray*}
}
and make use of the result (see e.g. Theorem 3.2.4 of Mardia, Kent \myand Bibby, 1979)
\begin{eqnarray*}
&&\left[
\begin{array}{c}      
s_2\\[1ex]
s_3
\end{array}
\right]
\Bigg| s_1\sim N\Bigg((s_1/\Vert\balpha_1\Vert^2)\left[
\begin{array}{c}      
\balpha_1^T\balpha_2\\[1ex]
\balpha_1^T\balpha_3
\end{array}
\right],\\[1ex]
&&\quad\qquad\qquad\qquad\left[
\begin{array}{ccc}      
\Vert\balpha_2\Vert^2 &\ \ \ &\balpha_2^T\balpha_3\\[1ex]
\balpha_2^T\balpha_3  & &\Vert\balpha_3\Vert^2
\end{array}
\right]  
-(1/\Vert\balpha_1\Vert^2)\left[
\begin{array}{ccc}      
(\balpha_1^T\balpha_2)^2 &\ \ \ & (\balpha_1^T\balpha_2)(\balpha_1^T\balpha_3)\\[1ex]
(\balpha_1^T\balpha_2)(\balpha_1^T\balpha_3) &\ \ \ & (\balpha_1^T\balpha_3)^2
\end{array}
\right]      
\Bigg).   
\end{eqnarray*}
Result (\ref{eq:LemmaOneC}) then follows via simple algebraic manipulations.

\jump
{\bf Lemma 2.} \textit{For all $a\in\real$ and $d\times1$ vectors $\bb$}

\setlength\arraycolsep{1pt}{
\begin{eqnarray}
\int_{\real^d}\Phi(a+\bb^T\bx)\phi_{\bI}(\bx)\,dx&=&\Phi\left(\frac{a}{\sqrt{\bb^T\bb+1}}\right),
\label{eq:LemmaTwoA}
\\[2ex]
\int_{\real^d}\,\bx\,\Phi(a+\bb^T\bx)\phi_{\bI}(\bx)\,dx&=&\frac{\bb}{\sqrt{\bb^T\bb+1}}\,
\phi\left(\frac{a}{\sqrt{\bb^T\bb+1}}\right)\quad\mbox{\textit{and}}
\label{eq:LemmaTwoB}
\\[2ex]
\int_{\real^d}\,\bx\bx^T\,\Phi(a+\bb^T\bx)\phi_I(\bx)\,dx&=&
\Phi\left(\frac{a}{\sqrt{\bb^T\bb+1}}\right)\!\bI
-\!\frac{a\bb\bb^T}{\sqrt{(\bb^T\bb+1)^3}}\phi\left(\frac{a}{\sqrt{\bb^T\bb+1}}\right).
\label{eq:LemmaTwoC}
\end{eqnarray}
}

\jump
{\bf Proof of Lemma 2.}

\vskip2mm
Suppose that $Z_1$ and $Z_2$ are independent $N(0,1)$ random variables. 
As defined in Section \ref{sec:fertility}, for a logical proposition $\Psc$, let  $I(\Psc)=1$ if $\Psc$ 
is true and $I(\Psc)=0$ if $\Psc$ is false. Then
$$P(Z_1\le a+\Vert\bb\Vert Z_2)=E\{I(Z_1\le a+\Vert\bb\Vert Z_2)\}=E[E\{I(Z_1\le a+\Vert\bb\Vert Z_2)|\bZ_2\}]=E\{h(\bZ_2)\}$$
where $h(z_2)\equiv E\{I(Z_1\le a+\Vert\bb\Vert Z_2)|Z_2=z_2\}$. But note that
$$h(z_2)=P(Z_1\le a+\Vert\bb\Vert z_2)=\Phi(a+\Vert\bb\Vert z_2)$$
which implies that 
$$P(Z_1\le a+\Vert\bb\Vert Z_2)=E\{\Phi(a+\Vert\bb\Vert Z_2)\}=\infint\Phi(a+\Vert\bb\Vert z)\phi(z)\,dz.$$
Then
$$P(Z_1\le a+\Vert\bb\Vert Z_2)=P(Z_1-\Vert\bb\Vert Z_2\le a)=P(X_3\le a)$$
where $X_3\equiv Z_1-\Vert\bb\Vert Z_2\sim N(0,1+\bb^T\bb)$ by independence of $Z_1$ and $Z_2$.
Then (\ref{eq:LemmaTwoA}) follows immediately.

Next, let $\be_i$ denote the $d\times1$ vector with $i$th entry equal to $1$ and with 
zeroes elsewhere. Then, using Lemma 1,the $i$th entry of the right-hand side of 
(\ref{eq:LemmaTwoB}) is
\setlength\arraycolsep{1pt}{
\begin{eqnarray*}
\int_{\real^d}\,(\be_i^T\bx)\,\Phi(a+\bb^T\bx)\phi_{\bI}(\bx)\,d\bx
&=&\{(\be_i^T\bb)/\Vert\bb\Vert\}\infint z\Phi(a+\Vert\bb\Vert z)\phi(z)\,dz\\[1ex]
&=&\,-\{(\be_i^T\bb)/\Vert\bb\Vert\}\infint \Phi(a+\Vert\bb\Vert z)\phi'(z)\,dz\\[1ex]
&=&\,(\be_i^T\bb)\infint \phi(a+\Vert\bb\Vert z)\phi(z)\,dz\\[1ex]
\end{eqnarray*}
}
where the last result follows via integration by parts. The last integrand is
$$(2\pi)^{-1}\exp\{-\smhalf(a+\Vert\bb\Vert z)^2-\smhalf z^2\}
=\phi\left(\frac{a}{\sqrt{\bb^T\bb+1}}\right)\phi\left(\frac{z+a\Vert\bb\Vert/(1+\bb^T\bb)}
{1\big/\sqrt{\bb^T\bb+1}}\right)
$$
and (\ref{eq:LemmaTwoB}) is an immediate consequence.

Lastly, because of (\ref{eq:LemmaOneC}), the $(i,j)$ entry of the right-hand side 
of (\ref{eq:LemmaTwoC}) is
\setlength\arraycolsep{1pt}{
\begin{eqnarray*}
&&\int_{\real^d}\,(\be_i^T\bx)(\be_j^T\bx)\,\Phi(a+\bb^T\bx)\phi_{\bI}(\bx)\,d\bx\\[1ex]
&&\qquad=(\be_i^T\be_j)\infint \Phi(a+\Vert\bb\Vert\,z)\phi(z)\,dz
+\frac{(\be_i^T\bb)(\be_j^T\bb)}{\Vert\bb\Vert^2}
\infint\Phi(a+\Vert\bb\Vert\,z)\phi''(z)\,dz\\[1ex]
&&\qquad=(\be_i^T\be_j)\Phi\left(\frac{a}{\sqrt{\bb^T\bb+1}}\right)
-\frac{(\be_i^T\bb)(\be_j^T\bb)}{\Vert\bb\Vert}
\infint\phi(a+\Vert\bb\Vert\,z)\phi'(z)\,dz\\[1ex]
\end{eqnarray*}
}
where the last result follows via integration by parts. The last integrand is
$$
-(2\pi)^{-1}z\exp\{-\smhalf(a+\Vert\bb\Vert z)^2-\smhalf z^2\}
=-z\phi\left(\frac{a}{\sqrt{\bb^T\bb+1}}\right)\phi\left(\frac{z+a\Vert\bb\Vert/(1+\bb^T\bb)}
{1\big/\sqrt{\bb^T\bb+1}}\right)
$$
and (\ref{eq:LemmaTwoC}) follows.

\rightline{\endproof}
\jump\jump

Next, we note a key connection between Kullback-Leibler projection onto
the unnormalized and normalized Multivariate Normal families. 
For the latter, we introduce the notation
$$\mbox{proj}_{\mbox{\tiny N}}[p](\bx)=q(\bx)$$
where $q$ is the Multivariate Normal density function minimizes $\mbox{KL}(p\Vert q)$.

\jump
{\bf Lemma 3.} \textit{Let $f\in L_1(\real^d)$ be such that $f\ge0$ and define
$C_f\equiv\int_{\real^d}f(\bx)\,d\bx$. Then}
$$\proj[f](\bx)=C_f\,\mbox{proj}_{\mbox{\tiny N}}[f/C_f](\bx).$$

\jump
{\bf Proof Lemma 3.}

Let $g(\cdot;\bdeta)$ be a generic unnormalized Multivariate Normal density function
with natural parameter vector $\bdeta$:
$$g(\bx;\bdeta)=\exp\left\{
\left[  
\begin{array}{c}
1\\
\bx\\
\vech(\bx\bx^T)
\end{array}
\right]^T\bdeta   
\right\}.
$$
Then the Kullback-Leibler divergence of $g(\cdot;\bdeta)$ from $f$ is 
$$\mbox{KL}\big(f\Vert g(\cdot;\bdeta)\big)=\int_{\real^d}
\big[f(\bx)\log\{f(\bx)/g(\bx;\bdeta)\}+g(\bx;\bdeta)-f(\bx)\big]\,d\bx
=\Ksc(\bdeta)+\mbox{const}
$$
where `const' denotes terms not depending on $\bdeta$ and
$$\Ksc(\bdeta)\equiv(2\pi)^{d/2}\exp\{\bdeta_0+A_N(\bdeta_{-0})\}-
\left[  
\begin{array}{c}
\int_{\real^d}f(\bx)\,d\bx\\[1ex]
\int_{\real^d}\bx\,f(\bx)\,d\bx\\[1ex]
\int_{\real^d}\vech(\bx\bx^T)\,f(\bx)\,d\bx
\end{array}
\right]^T\bdeta.
$$
The derivative vector of $\Ksc(\bdeta)$ is 
$$\Diff\Ksc(\bdeta)=(2\pi)^{d/2}\exp\{\bdeta_0+A_N(\bdeta_{-0})\}
\left[  
\begin{array}{c}
1\\[1ex]
\Diff\,A(\bdeta_{-0})^T
\end{array}
\right]^T
-
\left[  
\begin{array}{c}
\int_{\real^d}f(\bx)\,d\bx\\[1ex]
\int_{\real^d}\bx\,f(\bx)\,d\bx\\[1ex]
\int_{\real^d}\vech(\bx\bx^T)\,f(\bx)\,d\bx
\end{array}
\right]^T
$$
so the stationary condition, $\Diff\Ksc(\bdeta)^T=\bzero$, 
for the minimization of $\mbox{KL}\big(f\Vert g(\cdot;\bdeta)\big)$ is 
\begin{equation}
(2\pi)^{d/2}\exp\{\bdeta_0+A_N(\bdeta_{-0})\}
\left[  
\begin{array}{c}
1\\[1ex]
\nabla A_N(\bdeta_{-0})
\end{array}
\right]
=
\left[  
\begin{array}{c}
\int_{\real^d}f(\bx)\,d\bx\\[1ex]
\int_{\real^d}\bx\,f(\bx)\,d\bx\\[1ex]
\int_{\real^d}\vech(\bx\bx^T)\,f(\bx)\,d\bx
\end{array}
\right].
\label{eq:stationCond}
\end{equation}
with $\nabla A_N(\bdeta_{-0})\equiv\Diff\,A(\bdeta_{-0})^T$ denoting
the \emph{gradient} vector of $A(\bdeta_{-0})$.
It is easily checked that (\ref{eq:stationCond}) is satisfied by 
\begin{equation}
\setlength\arraycolsep{2pt}{
\begin{array}{rcl}
(\bdeta^*)_0&=&\log(C_f)-A_N(\bdeta^*_{-0})-\smhalf\,d\log(2\pi)\\[2ex]
\mbox{where}\quad \bdeta^*_{-0}&=&(\nabla A_N)^{-1}\left(\left[  
\begin{array}{c}
\int_{\real^d}\bx\,\{f(\bx)/C_f\}\,d\bx\\[2ex]
\int_{\real^d}\vech(\bx\bx^T)\,\{f(\bx)/C_f\}\,d\bx
\end{array}
\right]     
\right)
\end{array}
}
\label{eq:bdetaStar}
\end{equation}
with existence and uniqueness of $(\nabla A_N)^{-1}$ being guaranteed
by Proposition 3.2 of Wainwright \myand Jordan (2008). The Hessian matrix of 
$\Ksc(\bdeta)$ is 
$$\Hess\,\Ksc(\bdeta)=(2\pi)^{d/2}e^{\bdeta_0+A_N(\bdeta_{-0})}
\left\{
\left[  
\begin{array}{c}
1\\[1ex]
\nabla A_N(\bdeta_{-0})
\end{array}
\right]\left[  
\begin{array}{c}
1\\[1ex]
\nabla A_N(\bdeta_{-0})
\end{array}
\right]^T
+
\left[
\begin{array}{cc}
0&\bzero^T\\
\bzero&\Hess A_N(\bdeta_{-0})
\end{array}
\right]
\right\}
$$
From Proposition 3.1 of Wainwright \myand Jordan (2008), $A_N$ is strictly convex
on its domain and therefore $\Hess A_N(\bdeta_{-0})$ is positive definite.
Hence $\Hess\,\Ksc(\bdeta)$ is positive definite for all $\bdeta$ and so
(\ref{eq:bdetaStar}) is the unique minimizer of $\mbox{KL}\big(f\Vert g(\cdot;\bdeta)\big)$.
Therefore,
$$\proj[f](\bx)=\exp\left\{
\left[  
\begin{array}{c}
1\\
\bx\\
\vech(\bx\bx^T)
\end{array}
\right]^T\bdeta^*   
\right\}
$$
where $\bdeta^*$ is as given by (\ref{eq:bdetaStar}). However, $\bdeta_{-0}^*$ is 
the same natural parameter vector that arises via projection of $f/C_f$
onto the family of Multivariate Normal density functions and so
$$\proj_N[f/C_f](\bx)=\exp\left\{\left[  
\begin{array}{c}
\bx\\
\vech(\bx\bx^T)
\end{array}
\right]^T\bdeta_{-0}^*-A_N(\bdeta_{-0}^*)\right\}(2\pi)^{-d/2}
$$
which immediately leads to Lemma 3.

\rightline{\endproof}
\jump\jump

The proof  of Theorem 1 involves transferral between the common $N(\bmu,\bSigma)$ parameters
of the $d$-variate Normal distribution and the natural parameters corresponding
to the sufficient statistics $\bx$ and $\vech(\bx\bx^T)$. The transformations
in each direction are 
\begin{equation}
\left\{
\setlength\arraycolsep{1pt}{
\begin{array}{rcl}  
\bdeta_1&=&\bSigma^{-1}\bmu\\[2ex]
\bdeta_2&=&-\smhalf\bD_d^T\vecof(\bSigma^{-1})
\end{array}
}
\right.
\qquad\mbox{and}\qquad
\left\{
\setlength\arraycolsep{1pt}{
\begin{array}{rcl}  
\bmu&=&-\smhalf\big\{\vecof^{-1}\Big(\bD_d^{+T}\bdeta_2\big)\Big\}^{-1}\bdeta_1\\[2ex]
\bSigma&=&-\smhalf\big\{\vecof^{-1}\Big(\bD_d^{+T}\bdeta_2\big)\Big\}^{-1}
\end{array}
}
\right.
\label{eq:forwBackTranfs}
\end{equation}
Recall the notation 
$$\bv^{\otimes\,k}\equiv\left\{
\begin{array}{lcl}
1&\mbox{for}&k=0\\
\bv&\mbox{for}&k=1\\
\bv\bv^T&\mbox{for}&k=2\\
\end{array}
\right.
$$
and consider Kullback-Leibler projection of $\fInput/C_{\fInput}$
onto the family of $d$-variate Normal density functions
where
$$
\fInput(\bx)\equiv\Phi(c_0+\bc_1^T\bx)\exp\left\{
\left[  
\begin{array}{c}
\bx\\[1ex]
\vech(\bx\bx^T)
\end{array}
\right]^T   
\left[  
\begin{array}{c}
\bdetaOneInput\\[1ex]
\bdetaTwoInput
\end{array}
\right]
\right\},
$$
and $C_{\fInput}\equiv\int_{\real^d}\fInput(\bx)\,d\bx$.
Then the projection has mean and covariance matrix
\begin{equation}
\bmu^*=\Msc_1/\Msc_0\quad\mbox{and}\quad
\bSigma^*=\Msc_2/\Msc_0-(\Msc_1/\Msc_0)(\Msc_1/\Msc_0)^T
\label{eq:commonStar}
\end{equation}
where
$$\Msc_k\equiv 
\int_{\real^d}\bx^{\otimes\,k}\Phi(c_0+\bc_1^T\,\bx)
\exp\left\{
\left[  
\begin{array}{c}
\bx\\[1ex]
\vech(\bx\bx^T)
\end{array}
\right]^T   
\left[  
\begin{array}{c}
\bdetaOneInput\\[1ex]
\bdetaTwoInput
\end{array}
\right]
\right\}
\,d\bx.$$
Letting 
$$\bSigmaInput\equiv\,
-\smhalf\left\{\vecof^{-1}(\bD_{\dR}^{+T}\bdetaTwoInput) \right\}^{-1}
\quad\mbox{and}\quad\bmuInput\equiv\bSigmaInput\bdetaOneInput$$
be the common parameters corresponding to $\bdetaInput$ and making the 
change of variable $\bz=(\bSigmaInput)^{-1/2}(\bx-\bmuInput)$ 
we obtain
$$\Msc_k=(2\pi)^{d/2}e^{A_N(\bdetaInput)}\,\int_{\real^d}
\big(\bmuInput+(\bSigmaInput)^{1/2}\bz\big)^{\otimes\,k}
\Phi\Big((c_0+\bc_1^T\bmuInput)+\{(\bSigmaInput)^{1/2}\bc_1\}^T\bz\Big)\phi_{\bI}(\bz)\,d\bz.
$$
Lemma 2 and simple algebraic manipulations then give 
$$\Msc_1/\Msc_0=
\bmuInput+\frac{\bSigmaInput\bc_1\zeta'(r_2)}{\sqrt{\bc_1^T\bSigmaInput\bc_1+1}}.$$
and
\begin{eqnarray*}
\Msc_2/\Msc_0&=&\bmuInput(\bmuInput)^T
+\frac{\{\bSigmaInput\bc_1(\bmuInput)^T+\bmuInput\bc_1^T\bSigmaInput\}\zeta'(r_2)}{\sqrt{\bc_1^T\bSigmaInput\bc_1+1}}\\
&&\quad\qquad\qquad\qquad+\bSigmaInput-\frac{r\zeta'(r_2)\bSigmaInput
\bc_1\bc_1^T\bSigmaInput}{\bc_1^T\bSigmaInput\bc_1+1}.
\end{eqnarray*}
where
$$r_2\equiv\frac{2c_0-\bc_1^T\big\{\vecof^{-1}\big(\bD_{\dR}^{+T}\bdetaTwoInput\big)\big\}^{-1}\bdetaOneInput}
{\sqrt{2\left[2-\bc_1^T\big\{\vecof^{-1}\big(\bD_{\dR}^{+T}\bdetaTwoInput\big)\big\}^{-1}\bc_1\right]}}.$$
Combining these last two results and noting (\ref{eq:commonStar}) we obtain
the common parameter solutions
\setlength\arraycolsep{2pt}{
\begin{eqnarray*}
\bmu^*&=&\bmuInput+\frac{\bSigmaInput\bc_1\zeta'(r_2)}{\sqrt{\bc_1^T\bSigmaInput\bc_1+1}}\\
[1ex]
\bSigma^*&=&\bSigmaInput+\left\{\frac{\zeta''(r_2)}{\bc_1^T\bSigmaInput\bc_1+1}\right\}
\bSigmaInput\bc_1\bc_1^T\bSigmaInput.
\end{eqnarray*}
}
Transferral to natural parameters via (\ref{eq:forwBackTranfs}) and some
simple manipulations then lead to 
$$\left[
\begin{array}{c}
\bdeta_1^*\\
\bdeta_2^*
\end{array}
\right]=\Kprobit\left(\left[
\begin{array}{c}
\bdetaOneInput\\
\bdetaTwoInput
\end{array}
\right];c_0,\bc_1\right).
$$
Finally, 
\setlength\arraycolsep{1pt}{
\begin{eqnarray*}
\eta_0^*&=&\log(C_{\fInput})-\log\int_{\real^{\dR}}
\exp\left\{
\left[  
\begin{array}{c}
\bx\\[1ex]
\vech(\bx\bx^T)
\end{array}
\right]^T   
\left[  
\begin{array}{c}
\bdeta_1^*\\[1ex]
\bdeta_2^*
\end{array}
\right]
\right\}\,d\bx\\[1ex]
&=&\log(\Msc_0)-\smhalf\,\dR\log(2\pi)-A_N(\bdeta^*)
=\log\Phi(r_2)+A_N(\bdetaInput)-A_N(\bdeta^*)\\[1ex]
&=&\Cprobit\left(\left[
\begin{array}{c}
\bdetaOneInput\\
\bdetaTwoInput
\end{array}
\right],
\left[
\begin{array}{c}
\bdeta_1^*\\
\bdeta_2^*
\end{array}
\right]
;c_0,\bc_1\right)
\end{eqnarray*}
}

\section{Derivation of Algorithm \ref{alg:mainAlgo}}\label{sec:mainAlgoDerivn}

We now provide full justification of Algorithm \ref{alg:mainAlgo},
starting with a derivation of the message passing representation
used in Algorithm \ref{alg:mainAlgo}.

\subsection{Message Passing Representation Derivation}

The derivation of the message passing representation is based on
the infrastructure and results laid out in Minka (2005). The
treatment given there is for a generalization of Kullback-Leibler
divergence, known as \emph{$\alpha$-divergence}, and for approximation
of (normalized) density functions rather than general non-negative
$L_1$ functions. The Kullback-Leibler divergence minimization
problem given by (\ref{eq:KLijth}) corresponds to $\alpha=1$ in
the notation of Minka (2005). Following Section 4.1 of Minka (2005)
we then define the messages passed from the factors neighboring 
$\bu_i$ in Figure \ref{fig:frqMixModFacGraph} to be 
\begin{equation}
\mSUBpyijTOui\equiv\pEP(y_{ij}|\bu_i;\bbeta)\quad\mbox{and}\quad
\mSUBpuiTOui\equiv p(\bu_i;\bSigma).
\label{eq:msgDfn}
\end{equation}
Then, (54) of Minka (2005) invokes the definition
\begin{equation}
\mSUBuiTOpyij\equiv \mSUBpuiTOui\prod_{j'\ne j}
\mSUBpyijdTOui.
\label{eq:SNtoF}
\end{equation}

Result (60) of Minka (2005) with $\alpha=1$, $s'=1$
(since we are working with unnormalized rather than 
normalized Kullback-Leibler divergence) and the simplification
that there is only one stochastic node, namely $\bu_i$, 
provides the main factor to stochastic node message passing updates:
\begin{equation}
\mSUBpyijTOui\thickarrow
\frac{\proj\big[\mSUBuiTOpyij\,p(y_{ij}|\bu_i;\bbeta)\big](\bu_i)}
{\mSUBuiTOpyij},\ 1\le j\le n_i.
\label{eq:facToStochNodeForProbit}
\end{equation}
The other factor to stochastic node message passing update is,
trivially from (\ref{eq:msgDfn}),
$$\mSUBpuiTOui\thickarrow p(\bu_i;\bSigma).$$
The stochastic node to factor updates are, from (\ref{eq:SNtoF}),
$$\mSUBuiTOpyij\thickarrow \mSUBpuiTOui\prod_{j'\ne j}
\mSUBpyijdTOui,\quad 1\le j\le n_i.
$$
Next, we simplify these message updates to a programmable form.

\subsection{Simplification of the $\mSUBpyijTOui$ Updates}

From (\ref{eq:SNtoF}) it is apparent that $\mSUBpyijTOui$ is
an unnormalized Multivariate Normal density function and
therefore
$$
\mSUBuiTOpyij=\exp\left\{\left[\begin{array}{c}  
1\\
\bu_i\\
\vech(\bu_i\bu_i^T)
\end{array}
\right]^T\bdetaSUBuiTOpyij
\right\}
$$
with natural parameter vector $\bdetaSUBuiTOpyij$.
Introducing the abbreviation:
$$\bdetaSUBuiTOpyijABV\equiv\bdetaSUBuiTOpyij$$
we have
$$
\mSUBuiTOpyij=
\exp\big(\eta_0^{\otimes}\big)
\exp\left\{
\left[
\begin{array}{c}  
\bu_i\\
\vech(\bu_i\bu_i^T)
\end{array}
\right]^T\bdeta^{\otimes}_{-0}\right\}
$$
where $\eta_0^{\otimes}$ denotes the first entry of $\bdetaSUBuiTOpyijABV$
and $\bdeta^{\otimes}_{-0}$ contains the remaining entries.
Substitution info (\ref{eq:facToStochNodeForProbit}) leads to 
\begin{eqnarray*}
\mSUBpyijTOui&\thickarrow&
\frac{\proj\left[\exp\big(\eta_0^{\otimes}\big)
\exp\left\{
\left[
\begin{array}{c}  
\bu_i\\
\vech(\bu_i\bu_i^T)
\end{array}
\right]^T\bdeta^{\otimes}_{-0}\right\}
\,\Phi\big((2y_{ij}-1)(\bbeta^T\bxF_{ij}+\bu_i^T\bxR_{ij})\big)\right](\bu_i)}
{\exp\big(\eta_0^{\otimes}\big)
\exp\left\{
\left[
\begin{array}{c}  
\bu_i\\
\vech(\bu_i\bu_i^T)
\end{array}
\right]^T\bdeta^{\otimes}_{-0}\right\}}\\[2ex]
&=&
\frac{\mbox{proj}\left[
\,\Phi\big(c_{0,ij}+\bc_{1,ij}^T\,\bu_i\big)\exp\left\{
\left[
\begin{array}{c}  
\bu_i\\
\vech(\bu_i\bu_i^T)
\end{array}
\right]^T\bdeta^{\otimes}_{-0}\right\}\right](\bu_i)}
{\exp\left\{
\left[
\begin{array}{c}  
\bu_i\\
\vech(\bu_i\bu_i^T)
\end{array}
\right]^T\bdeta^{\otimes}_{-0}\right\}}
\end{eqnarray*}
where
$$c_{0,ij}\equiv(2y_{ij}-1)(\bbeta^T\bxF_{ij})\quad\mbox{and}\quad
\bc_{1,ij}\equiv (2y_{ij}-1)\bxR_{ij}.
$$
Using Theorem 1:
$$\mSUBpyijTOui\leftarrow
\exp\left\{\left[\begin{array}{c}  
1\\
\bu_i\\
\vech(\bu_i\bu_i^T)
\end{array}
\right]^T 
\bdetaSUBpyijTOui
\right\}
$$
where the linear and quadratic coefficient updates are
\begin{eqnarray*}
\big(\bdetaSUBpyijTOui\big)_{-0}&\thickarrow& \Kprobit\Big(\big(\bdetaSUBuiTOpyij\big)_{-0};
(2y_{ij}-1)(\bbeta^T\bxF_{ij}),(2y_{ij}-1)\bxR_{ij}\Big)\\
&&\qquad\qquad-\big(\bdetaSUBuiTOpyij\big)_{-0}
\end{eqnarray*}
and the constant coefficient update is
\begin{eqnarray*}
\big(\bdetaSUBpyijTOui\big)_0&\thickarrow& \Cprobit\Big(\big(\bdetaSUBuiTOpyij\big)_{-0},
\big(\bdetaSUBpyijTOui\big)_{-0}\\
&&\qquad\qquad+\big(\bdetaSUBuiTOpyij\big)_{-0};(2y_{ij}-1)(\bbeta^T\bxF_{ij}),(2y_{ij}-1)\bxR_{ij}\Big).
\end{eqnarray*}

\subsection{Simplification of the $\mSUBpuiTOui$ Update}

The second definition in (\ref{eq:msgDfn}) gives
$$\mSUBpuiTOui\thickarrow p(\bu_i;\bSigma)=\exp\left\{\left[\begin{array}{c}  
1\\
\bu_i\\
\vech(\bu_i\bu_i^T)
\end{array}
\right]^T 
\bdetaSUBSigma
\right\}.
$$
Therefore, if $\bdetaSUBpuiTOui$ denotes the natural parameter vector 
of $\mSUBpuiTOui$ then it has the trivial update
$$\bdetaSUBpuiTOui\thickarrow \bdetaSUBSigma.$$

\subsection{Simplification of the $\mSUBuiTOpyij$ Updates}

Given the simplified forms of the messages in the two previous 
subsections we have from (\ref{eq:SNtoF}):
\begin{eqnarray*}
\mSUBuiTOpyij&\thickarrow&\exp\left\{\left[\begin{array}{c}  
1\\
\bu_i\\
\vech(\bu_i\bu_i^T)
\end{array}
\right]^T\bdetaSUBpuiTOui
\right\}\\[1ex]
&&\qquad\times\prod_{j'\ne j}
\exp\left\{\left[\begin{array}{c}  
1\\
\bu_i\\
\vech(\bu_i\bu_i^T)
\end{array}
\right]^T\bdetaSUBpyijdTOui
\right\}\\[1ex]
\end{eqnarray*}
which leads to 
{\setlength\arraycolsep{2pt}
\begin{eqnarray*}
\bdetaSUBuiTOpyij&\thickarrow&\bdetaSUBpuiTOui+\sum_{j'\ne j}\bdetaSUBpyijdTOui\\[1ex]
&=&\bdetaSUBpuiTOui+\mbox{SUM}\{\etaSUBpyiTOui\}-\bdetaSUBpyijTOui.
\end{eqnarray*}
}
\subsection{Assembly of All Natural Parameter Updates}

We now return to the message passing protocol given in Section \ref{sec:messPassFormul}:

\begin{minipage}[t]{132mm}
\hrule
\vskip3mm
\begin{itemize}
\item[] Initialize all factor to stochastic node messages.
\item[] Cycle until all factor to stochastic node messages converge:
\begin{itemize}
\item[]For each factor:
\begin{itemize}
\item[] Compute the messages passed to the factor using (\ref{eq:StoFI})
or (\ref{eq:StoFII}).
\item[] Compute the messages passed from the factor using (\ref{eq:FtoSI})
or (\ref{eq:FtoSII}).
\end{itemize}
\end{itemize}
\end{itemize}
\hrule
\end{minipage}

\vskip3mm\noindent
For the factors $p(y_{ij}|\bu_i;\bbeta)$: 
\begin{itemize}
\item[] computing the messages passed to each of these factors reduces to
\item[] $\bdetaSUBuiTOpyij\thickarrow\bdetaSUBpuiTOui+\mbox{SUM}\{\etaSUBpyiTOui\}-\bdetaSUBpyijTOui$
\item[] and computing the messages passed from these factors reduces to 
\item[] $\Big(\etaSUBpyijTOui\Big)_{-0}\thickarrow\Kprobit\Big(\big(\etaSUBuiTOpyij\big)_{-0};c_{0,ij},\bc_{1,ij}\Big)$
\item[] $\qquad\qquad\qquad\qquad\qquad\qquad\qquad\qquad\qquad-\big(\etaSUBuiTOpyij\big)_{-0}$\\
\item[] and 
\item[] $\Big(\etaSUBpyijTOui\Big)_0\thickarrow\Cprobit\Big(\big(\etaSUBuiTOpyij\big)_{-0},
\big(\etaSUBpyijTOui\big)_{-0}$
\item[]$\qquad\qquad\qquad\qquad\qquad\qquad\qquad\quad\qquad
+\big(\etaSUBuiTOpyij\big)_{-0};c_{0,ij},\bc_{1,ij}\Big)$.
\end{itemize}

\vskip3mm\noindent
For the factors $p(\bu_i;\bSigma)$: 
\begin{itemize}
\item[] computing the messages passed from these factors reduces to 
\item[] $\bdetaSUBuiTOpui\thickarrow\displaystyle{\sum_{j=1}^{{n_i}}}\,\bdetaSUBpyijTOui$
\item[] and computing the messages passed to these factors reduces to 
\item[] $\bdetaSUBpuiTOui\to\biggerbdeta_{\bSigma}$.
\end{itemize}

Algorithm \ref{alg:mainAlgo} is essentially these natural parameter updates being
cycled until convergence. The update for $\Big(\etaSUBpyijTOui\Big)_0$ can be 
moved outside of the cycle loop without affecting convergence. 
Also, the $\bdetaSUBuiTOpui$ updates are redundant and are omitted from
Algorithm \ref{alg:mainAlgo}.

\section{Derivation of Starting Values Recommendation}\label{sec:sttValsDeriv}

We now derive useful starting values for the $\etaSUBpyijTOui$ that
have to be initialized in Algorithm \ref{alg:mainAlgo}.
Note that 
$$\log\,p(y_{ij}|\bu_i;\bbeta)=\sum_{j=1}^{n_i}\{\zeta(a_{ij})-\log(2)\}\quad\mbox{where}\quad
a_{ij}\equiv(2y_{ij}-1)(\bbeta^T\bxF_{ij}+\bu_i^T\bxR_{ij})$$
and $\zeta$ is as defined in Section \ref{sec:projUNN}.
Let $\buhat_i$ be a prediction of $\bu_i$ and consider the following
expansion of the data-dependent component of $\ell(\bbeta,\bSigma)$:
\begin{eqnarray*}
\zeta(a_{ij})
&=&\zeta\big(\ahat_{ij}+(\bu_i-\buhat_i)^T\bxR_{ij}(2y_{ij}-1)\big)\\[1ex]
&=&\zeta(\ahat_{ij})+(\bu_i-\buhat_i)^T\bxR_{ij}y_{ij}^{\dagger}
\zeta'\big(\ahat_{ij}\big)+\smhalf\{(\bu_i-\buhat_i)^T\bxR_{ij}(2y_{ij}-1)\}^2
\zeta''(\ahat_{ij})+\ldots\\[2ex]
&=&\left[
\begin{array}{c}
1\\
\bu_i-\buhat_i\\
\vech\big((\bu_i-\buhat_i)(\bu_i-\buhat_i)^T\big)
\end{array}
\right]^T
{\check\bdeta}_{ij}+\ldots
\end{eqnarray*}
where, as in Section \ref{sec:sttVals}, 
$\ahat_{ij}\equiv(2y_{ij}-1)(\bbeta^T\bxF_{ij}+\bu_i^T\bxR_{ij})$,
and
$${\check\bdeta}_{ij}\equiv\left[
\begin{array}{c}
\zeta(\ahat_{ij})\\[1.5ex]
\bxR_{ij}(2y_{ij}-1)
\zeta'(\ahat_{ij})\\[1.5ex]
\smhalf\zeta''(\ahat_{ij})
\bD_{\dR}^T\vecof\big(\bxR_{ij}(\bxR_{ij})^T\big)
\end{array}
\right].
$$
It follows that the quadratic approximation to
$\log\,p(y_{ij}|\bu_i;\bbeta)$ based on Taylor expansion
about $\buhat_i$ is $\log\,\pcheck(y_{ij}|\bu_i;\bbeta)$ 
where 
$$\pcheck(y_{ij}|\bu_i;\bbeta)\equiv
\exp\left\{\left[
\begin{array}{c}
1\\
\bu_i-\buhat_i\\
\vech\big((\bu_i-\buhat_i)(\bu_i-\buhat_i)^T\big)
\end{array}
\right]^T
{\check\bdeta}_{ij}
\right\}.
$$
The starting value recommendation for $\bdetaSUBpyijTOui$ is 
based on replacement of $\pcheck(y_{ij}|\bu_i;\bbeta)$ by $p(y_{ij}|\bu_i;\bbeta)$
in (\ref{eq:facToStochNodeForProbit}): 
$$
\mSUBpchkyijTOui\thickarrow
\frac{\proj[\mSUBuiTOpchkyij\,\pcheck(y_{ij}|u_i;\bbeta)](\bu_i)}
{\mSUBuiTOpchkyij}\\
=\pcheck(y_{ij}|\bu_i;\bbeta)
$$
with the $\proj[\cdot]$ being superfluous in this case due to
$\pcheck(y_{ij}|u_i;\bbeta)$ being already in the Multivariate Normal family.
The starting value for $\etaSUBpyijTOui$ that arises from this substitution 
is then given by
$$\exp\left\{\left[
\begin{array}{c}
1\\
\bu_i\\
\vech\big(\bu_i\bu_i^T\big)
\end{array}
\right]^T
\etaSUBpyijTOui^{\mbox{\tiny start}}\right\}
=
\exp\left\{\left[
\begin{array}{c}
1\\
\bu_i-\buhat_i\\
\vech\big((\bu_i-\buhat_i)(\bu_i-\buhat_i)^T\big)
\end{array}
\right]^T
{\check\bdeta}_{ij}
\right\}.
$$
By matching coefficients of like terms we arrive at 
$$\etaSUBpyijTOui^{\mbox{\tiny start}}
=\left[
\begin{array}{c}
\eta_0^{\mbox{\tiny start}}\\[0.5ex]
(2y_{ij}-1)\zeta'(\ahat_{ij})\bxR_{ij}-\zeta''(\ahat_{ij})\bxR_{ij}(\bxR_{ij})^T\buhat_i\\[2ex]
\smhalf\zeta''(\ahat_{ij})\bD_{\dR}^T\vecof\big(\bxR_{ij}(\bxR_{ij})^T\big)
\end{array}
\right]
$$
where 
$$\eta_0^{\mbox{\tiny start}}=\zeta(\ahat_{ij})-
(2y_{ij}-1)\zeta'(\ahat_{ij})(\bxR_{ij})^T\buhat_i
+\smhalf\zeta''(\ahat_{ij})\{(\bxR_{ij})^T\buhat_i\}^2.
$$

In Algorithm \ref{alg:mainAlgo} the cycle loop corresponds to 
determination of the natural parameter vector
$$\Big(\etaSUBpyijTOui\Big)_{-0}$$
implying that the first entry of $\etaSUBpyijTOui^{\mbox{\tiny start}}$ is not
needed for these iterations. Hence, we can instead set $\eta_0^{\mbox{\tiny start}}=0$
without affecting Algorithm \ref{alg:mainAlgo}. We now have (\ref{eq:etaSttExpr}).

\section{Details of Confidence Interval Calculations}\label{sec:confIntDetails}

Here we provide full details of approximate confidence intervals calculations
based on quasi-Newton maximization of $\ellEP(\bbeta,\bSigma)$. The calculations
depend on the following ingredients:
\begin{itemize}
\item some additional convenient matrix notation.
\item formulae for transformation from the parameter vector 
$\btheta\equiv\vech\big(\smhalf\log(\bSigma)\big)$ to a
parameter vector $\bomega$ that is more appropriate for
confidence interval construction.
\item formulae for the reverse transformation: from $\bomega$ to 
$\btheta$.
\item a quasi-Newton optimization-based strategy for calculating
confidence intervals for the entries of $\bomega$, which are then
easily transformed to confidence intervals for interpretable covariance
matrix parameters, as illustrated in Figures \ref{fig:JASAsimStudy1CIplot}
and \ref{fig:JASAsimStudy2CIplot}.
\end{itemize}

\subsection{Additional Matrix Notation}

For a $d\times d$ matrix $\bA$ define $\diagonal(\bA)$ to be the
$d\times1$ vector consisting of the diagonal entries of $\bA$
and, provided $d\ge2$, define $\vecbd(\bA)$ to be the $\smhalf\,d(d-1)$ vector containing
the entries of $\bA$ that are below the diagonal of $\bA$ in
order from left to right and top to bottom. For example,
$$\mbox{diagonal}\left(   
\left[
\begin{array}{rrrr}
1 &\ \  5 &\ \ \ \ 9 &\ \   13\\
2 & 6 & 10 & 14\\
3 & 7 & 11 & 15\\
4 & 8 & 12 & 16
\end{array}
\right]
\right)
=
\left[
\begin{array}{c}
1\\
6\\
11\\
16
\end{array}
\right]
\quad\mbox{and}\quad
\mbox{vecbd}\left(   
\left[
\begin{array}{rrrr}
1 &\ \  5 &\ \ \ \ 9 &\ \   13\\
2 & 6 & 10 & 14\\
3 & 7 & 11 & 15\\
4 & 8 & 12 & 16
\end{array}
\right]
\right)
=
\left[
\begin{array}{r}
2\\
3\\
4\\
7\\
8\\
12
\end{array}
\right].
$$
In addition, if each of $\ba$ and $\bb$ are $d\times 1$ vectors then
$\ba\odot\bb$ is $d\times 1$ vector of element-wise products and
$\ba/\bb$ is $d\times 1$ vector of element-wise quotients.
Similarly, $\log(\ba)$ and $\tanh(\ba)$ are obtained in an
element-wise fashion.

\subsection{Transformation from $\btheta$ to $\bomega$}\label{sec:thetaTOomega}

Given a $\smhalf d(d+1)\times 1$ vector $\btheta$, 
its corresponding $\bomega$ vector of the same
length is found via the steps:

\begin{center}
\begin{minipage}[t]{130mm}
\begin{enumerate}
\item Obtain the spectral decomposition 
$\vech^{-1}(\btheta)=\bUbtheta\,\diag(\blambdabtheta)\bUbtheta^T$.
\item Set $\bSigma=\bUbtheta\diag\{\exp(2\blambdabtheta)\}\bUbtheta^T$.
\item 
\begin{enumerate}
\item If $d=1$ then $\bomega=\log(\sqrt{\bSigma})$.
\item If $d>1$ then
$$\bomega=\left[           
\begin{array}{c}
\log\Big(\sqrt{\mbox{diagonal}(\bSigma)}\Big)\\[2ex]
\tanh^{-1}\Bigg(\vecbd(\bSigma)\Big/
\sqrt{\mbox{vecbd}\Big(\mbox{diagonal}(\bSigma)\mbox{diagonal}(\bSigma)^T\Big)}\Bigg)
\end{array}
\right].
$$
\end{enumerate}
\end{enumerate}
\end{minipage}
\end{center}

\subsection{Transformation from $\bomega$ to $\btheta$}\label{sec:omegaTOtheta}

Given a $\smhalf d(d+1)\times 1$ vector $\bomega$, 
its corresponding $\btheta$ vector of the same
length is found via the steps:

\begin{center}
\begin{minipage}[t]{130mm}
\begin{enumerate}
\item Form the $d\times d$ symmetric matrix $\bSigma$ as follows:
\begin{enumerate}
\item If $d=1$ then $\bSigma=\exp(2\bomega)$.
\item If $d>1$ then let $\bomega_1$ denote the first $d$ entries
of $\bomega$ and $\bomega_2$ denote the remaining $\smhalf d(d-1)$ entries
of $\bomega$.
\begin{enumerate}
\item Set $\mbox{diagonal}(\bSigma)=\exp(2\bomega_1)$
\item Obtain the below-diagonal entries of $\bSigma$ so that
$$\vecbd(\bSigma)=\tanh(\bomega_2)\odot\vecbd\big(\exp(\bomega_1)\exp(\bomega_1)^T\big)$$
holds. Obtain the above-diagonal entries of $\bSigma$ such that symmetry of $\bSigma$ is enforced.
\end{enumerate}
\end{enumerate}
\item Obtain the spectral decomposition: $\bSigma=\bUbSigma\mbox{diag}(\blambdabSigma)\bUbSigma^T$.
\item Obtain $\btheta=\vech\Big(\smhalf\bUbSigma\mbox{diag}\{\log(\blambdabSigma)\}\bUbSigma^T\Big)$.
\end{enumerate}
\end{minipage}
\end{center}

\subsection{Quasi-Newton Optimization-Based Confidence Interval Calculations}

The steps for obtaining confidence intervals for each
of the interpretable parameters are:

\begin{center}
\begin{minipage}[t]{130mm}
\begin{enumerate}
\item Obtain $({\widehat\bbetaEP},{\widehat\bthetaEP})$ using a quasi-Newton
optimization routine applied the expectation propagation-approximate log-likelihood
$\ellEP$ with unconstrained input parameters $(\bbeta,\btheta)$.
\item Obtain ${\widehat\bomegaEP}$ corresponding to ${\widehat\bthetaEP}$ 
using the steps given in Section \ref{sec:thetaTOomega}.
\item Call the quasi-Newton optimization routine with input parameters
$(\bbeta,\bomega)$ instead of $(\bbeta,\btheta)$, and initial value
$({\widehat\bbetaEP},{\widehat\bomegaEP})$. In this call, request 
that the Hessian matrix $\Hess\ellEP(\bbeta,\bomega)$ at the 
maximum $({\widehat\bbetaEP},{\widehat\bomegaEP})$ be computed.
The steps given in Section \ref{sec:omegaTOtheta} are used to obtain the
corresponding $(\bbeta,\btheta)$ vector for evaluation of $\ellEP$ 
via the version of $\ellEP$ used in 1. for the optimization.
\item Form $100(1-\alpha)$\% confidence intervals for the entries of 
$(\bbeta,\bomega)$ using
$$\left[\begin{array}{c}       
{\widehat\bbetaEP}\\[2ex]
{\widehat\bomegaEP}
\end{array}
\right]\pm\Phi^{-1}(1-\smhalf\,\alpha)\sqrt{-\mbox{diagonal}\big(
\{\Hess\,\ellEP(\widehat{\bbetaEP},\widehat{\bomegaEP})\}^{-1}\big)}.
$$
\item Transform the confidence intervals limits for the $\bomega$ component,
using the functions $\exp$ and $\tanh$, to instead correspond to the standard deviation
and correlation parameters:
$$\left[           
\begin{array}{c}
\sqrt{\mbox{diagonal}(\bSigma)}\\[2ex]
\vecbd(\bSigma)\Big/
\sqrt{\mbox{vecbd}\Big(\mbox{diagonal}(\bSigma)\mbox{diagonal}(\bSigma)^T\Big)}
\end{array}
\right].
$$
\end{enumerate}
\end{minipage}
\end{center}

\section{Details of Approximate Best Prediction}\label{sec:bestPredDetails}

For the binary mixed model (\ref{eq:probitMixMod}), the best prediction of 
$\bu_i$ is 
\begin{eqnarray*}
\mbox{BP}(\bu_i)&=&E(\bu_i|\by)=E(\bu_i|\by_i)
=\int_{\real^{\dR}}\bu_i\,p(\bu_i|\by_i;\bbeta,\bSigma)\,d\bu_i\\
&=&\int_{\real^{\dR}}\bu_i\left\{\frac{p(\by_i|\bu_i;\bbeta)p(\bu_i;\bSigma)}
{\int_{\real^{\dR}}p(\by_i|\bu_i;\bbeta)p(\bu_i;\bSigma)}\right\}\,d\bu_i
\end{eqnarray*}
where $\by_i\equiv(y_{i1},\ldots,y_{in_i})$.  Now note that Algorithm 1 
involves replacement of 
$$p(\by_i|\bu_i;\bbeta)p(\bu_i;\bSigma)\quad\mbox{by}\quad 
\exp\left\{\left[\begin{array}{c}   
1 \\[0ex]
\bu_i\\[0ex]
\vech(\bu_i\bu_i^T)
\end{array}
\right]^T{\widehat\bdetaEP}_i\right\}
$$
where ${\widehat\bdetaEP}_i$ is defined by (\ref{eq:bdetaEPdfn}).
This leads to the approximation
\setlength\arraycolsep{1pt}{
\begin{eqnarray*}
\BPEP(\bu_i)&=&E({\widehat\buEP}_i)\ \mbox{where}\ {\widehat\buEP}_i\ \mbox{is Multivariate Normal with
natural parameter ${\widehat\bdetaEP}_i$}\\[1ex]
&=&-\smhalf\Big\{\vecof^{-1}\Big(
\bD_d^{+T}{\widehat\bdetaEP}_{i2}\Big)\Big\}^{-1}{\widehat\bdetaEP}_{i1}.
\end{eqnarray*}
}

Using (13.7) of McCulloch, Searle \myand Neuhaus (2008), the covariance matrix 
of $\mbox{BP}(\bu_i)-\bu_i$ is
$$\Cov\{\mbox{BP}(\bu_i)-\bu_i\}=E_{\by_i}\{\Cov(\bu_i|\by_i)\}.$$
The expectation propagation approximation of $\Cov(\bu_i|\by_i)$ is
$$\CovEP(\bu_i|\by)=-\smhalf\Big\{\vecof^{-1}
\Big(\bD_d^{+T}{\widehat\bdetaEP}_{i2}\Big)\Big\}^{-1}.$$
However, approximation of $\Cov\{\mbox{BP}(\bu_i)-\bu_i\}$ is hindered
by the expectation over the $\by_i$ vector.

\end{document}